# A Physics-Constrained Implicit Profile Network for Continuous Reconstruction of Tropical Cyclone Near-Surface Wind Profiles

**Jian Ma[1,2], Yilin Yang[1], Robert Rogers[1], Jun A. Zhang[3], Jie Tang[1,2]**

[1]Asia-Pacific Typhoon Collaborative Research Center, Shanghai, China

[2]Shanghai Typhoon Institute, CMA, Shanghai, China

[3]University of Miami, Miami, FL, USA

Corresponding author: Jie Tang (tangj@typhoon.org.cn)

**Key Points:**

- **We develop a physics-constrained implicit neural network Smartyphoon-BL3D for continuous reconstruction of tropical cyclone near-surface wind profiles from sparse dropsonde observations.**
- **The model integrates physics-based constraints with deep learning to improve profile realism while exactly preserving observed 10 m wind conditions.**
- **The framework accurately reconstructs boundary-layer wind profiles across diverse tropical cyclone environments and supports future AI-enabled Earth-system applications.**

**Abstract**

Accurate reconstruction of tropical cyclone near-surface wind profiles is essential for boundary-layer research, storm-surge prediction, offshore engineering, and marine operations. However, dropsonde observations are vertically irregular and spatially sparse, while conventional empirical profile models have limited capability to represent the complex, height-dependent wind structures observed in tropical cyclone boundary layers. We develop a Physics-Constrained Implicit Profile Network (Smartyphoon-BL3D) that reconstructs continuous radial and tangential wind profiles between 10 and 300 m from sparse dropsonde observations and compact storm-environmental information. The model combines a continuous implicit neural representation with physics-informed constraints that preserve observed 10 m winds, suppress unrealistic vertical oscillations, and promote physically consistent profile evolution. Evaluation using quality-controlled NOAA Hurricane Research Division dropsonde observations collected from 93 tropical cyclones during 2001–2022 demonstrates strong agreement with observations, yielding coefficients of determination ($R^2$) of 0.948 for radial wind, 0.963 for tangential wind, and 0.963 for wind-speed magnitude. Reconstruction errors increase with storm intensity, particularly for major hurricanes, reflecting the greater complexity of intense inner-core boundary-layer structures. A Hurricane Lee (2023) case study further demonstrates the framework's ability to extend satellite-derived surface winds into continuous three-dimensional near-surface wind fields while maintaining consistency with independent dropsonde observations. These results demonstrate the potential of physics-constrained implicit neural networks for reconstructing continuous atmospheric profiles from sparse observations and provide a foundation for future AI-enabled applications in tropical cyclone analysis, forecasting, and offshore hazard assessment.

**Plain Language Summary**

Near the ocean surface, tropical cyclone winds change rapidly with height, but direct measurements are limited because aircraft dropsondes provide only sparse and irregular observations. Continuous wind profiles are important for understanding hurricane boundary-layer processes, improving storm-surge prediction, supporting offshore engineering, and assessing coastal hazards. In this study, we developed an artificial intelligence model that combines machine learning with physical principles to reconstruct continuous wind profiles from sparse observations. The model is designed to preserve the observed surface winds while generating realistic changes in wind speed and direction with height. Tests using more than two decades of NOAA hurricane observations show that the method accurately reproduces the vertical structure of tropical cyclone winds over a wide range of storm intensities. The framework can also extend satellite-derived surface wind measurements into three-dimensional near-surface wind fields, providing new opportunities for hurricane research, operational forecasting, and engineering applications.

## 1 Introduction

ropical cyclones are among the most destructive atmospheric systems over the ocean. Their near-surface wind profiles govern air–sea momentum exchange, storm-surge forcing, wave generation, and the wind loads experienced by offshore platforms, wind turbines, bridges, ships, and coastal infrastructure (Powell et al., 2003; Vickery et al., 2009). Unlike a single 10 m wind speed, a vertical wind profile describes how radial inflow, tangential circulation, total wind speed, and wind direction vary with height, providing a direct link between tropical cyclone boundary-layer dynamics and engineering applications.Despite their importance, reliable characterization of tropical cyclone near-surface wind profiles remains challenging. Observations within the inner-core region are costly and difficult to obtain. Dropsondes have been used for decades to measure vertical profiles of wind, temperature, and moisture in tropical cyclones (Hock & Franklin, 1999; Aberson et al., 2023).

However, dropsonde observations are irregularly sampled in the vertical, sparse in storm-relative space, and unevenly distributed among storms and intensity categories (Uhlhorn et al., 2014). These characteristics make simple interpolation or case-by-case fitting inadequate for reconstructing continuous profiles at unobserved heights or locations.Classical logarithmic-law, power-law, Ekman-spiral, and empirical regression models (e.g., Holland, 1980) provide transparent and computationally efficient descriptions of the tropical cyclone boundary layer (Kepert, 2001; Smith & Montgomery, 2010; Zhang & Uhlhorn, 2012). However, their prescribed functional forms are often too restrictive to represent the complex, non-monotonic vertical structures observed under strong-wind conditions (Rogers et al., 2013; Ming et al., 2022). In particular, the transition layer between approximately 60 and 200 m often exhibits enhanced curvature, rapid flow turning, and height-dependent inflow-angle variations associated with turbulent friction, vertical shear, sea-state-dependent roughness, and storm asymmetry.

These limitations motivate more flexible approaches capable of representing realistic boundary-layer wind structures without prescribing a fixed functional form.Recent advances in machine learning provide new opportunities for reconstructing complex atmospheric structures from sparse observations. However, purely data-driven models may produce nonphysical oscillations, unstable extrapolation, or inconsistencies with observed surface conditions when physical constraints are absent. An effective reconstruction framework should therefore combine the flexibility of neural networks with physical constraints that preserve vertical continuity (Reichstein et al., 2019), maintain hemispheric wind-rotation consistency, represent realistic near-surface behavior, and remain fully consistent with the observed 10 m wind components. Recent studies have explored deep-learning approaches for tropical cyclone wind reconstruction, including physics-informed neural networks for sparse data assimilation (Eusebi et al., 2024) and adversarial learning for multi-level wind reconstruction from dropsonde observations (Han et al., 2025). These studies demonstrate the growing potential of integrating physical knowledge with machine learning for tropical cyclone boundary-layer applications.

To address these challenges, we develop a Physics-Constrained Implicit Profile Network (Smartyphoon-BL3D) for continuous reconstruction of tropical cyclone near-surface wind profiles from sparse dropsonde observations and storm-environmental information. The model represents radial and tangential wind components as continuous functions of height conditioned on storm-relative environmental variables. It combines an implicit neural representation with physics-informed constraints to preserve the observed 10 m wind components while promoting realistic vertical profile evolution and suppressing nonphysical artifacts. Training integrates a weighted Huber loss with physically motivated regularization terms that improve profile smoothness, hemispheric wind consistency, realistic inflow-angle behavior, and stable model optimization. Details of the network architecture and training procedure are presented in Section 3. **The main contributions of this study are summarized as follows:**

(1) A physics-constrained implicit neural network is developed for continuous reconstruction of tropical cyclone near-surface radial and tangential wind profiles.

(2) The proposed framework combines continuous neural representations with physically informed constraints to preserve observed surface winds while producing physically realistic vertical wind structures.

(3) The model is trained using a multi-term objective function that balances observational fidelity with physical consistency and stable optimization.

(4) The framework is comprehensively evaluated using quality-controlled NOAA HRD dropsonde observations, demonstrating accurate reconstruction across a wide range of tropical cyclone intensities and realistic representation of boundary-layer wind structures.

## 2 Data Description and Preprocessing

### 2.1 Dropsonde Data and Quality Control

This study uses quality-controlled tropical cyclone dropsonde observations collected by the NOAA Hurricane Research Division (HRD) during aircraft reconnaissance missions between 2001 and 2022. The dataset spans 93 tropical cyclones over the Atlantic and eastern North Pacific basins and includes storms covering a broad range of intensities, sizes, and environmental conditions. The original archive contains 13,058 records. Prior to model training and evaluation, all observations underwent the standard NOAA HRD quality-control procedures before analysis. Only records containing the required variables, including height, zonal and meridional wind components, storm center position, maximum storm wind speed, radius of maximum wind, and dropsonde relative location, were considered. The wind profiles were first converted to finite numerical arrays, and profile levels with missing height, zonal wind, or meridional wind values were removed. Duplicate height levels were merged by averaging the corresponding wind components. The analysis was then restricted to the near-surface layer from 10 to 300 m. This vertical domain focuses on the layer most relevant to offshore engineering applications and boundary-layer process studies; extension to higher altitudes (e.g., the full dropsonde depth of approximately 3000 m, corresponding to typical aircraft flight levels) is discussed as a future research direction in Section 5. The raw variables used in this study are summarized in Table 1.

**Table 1.** Variables in the raw data

| Name | Symbol | Unit |
|---|---|---|
| Cyclone center longitude | $tclon$ | degree |
| Cyclone center latitude | $tclat$ | degree |
| Radius of maximum wind | $tcrmw$ | m |
| Maximum wind speed | $tcwnd$ | m/s |
| Observation-point longitude | $lon$ | degree |
| Observation-point latitude | $lat$ | degree |
| Zonal wind at surface | $U_{10}$ | m/s |
| Meridional wind at surface | $V_{10}$ | m/s |

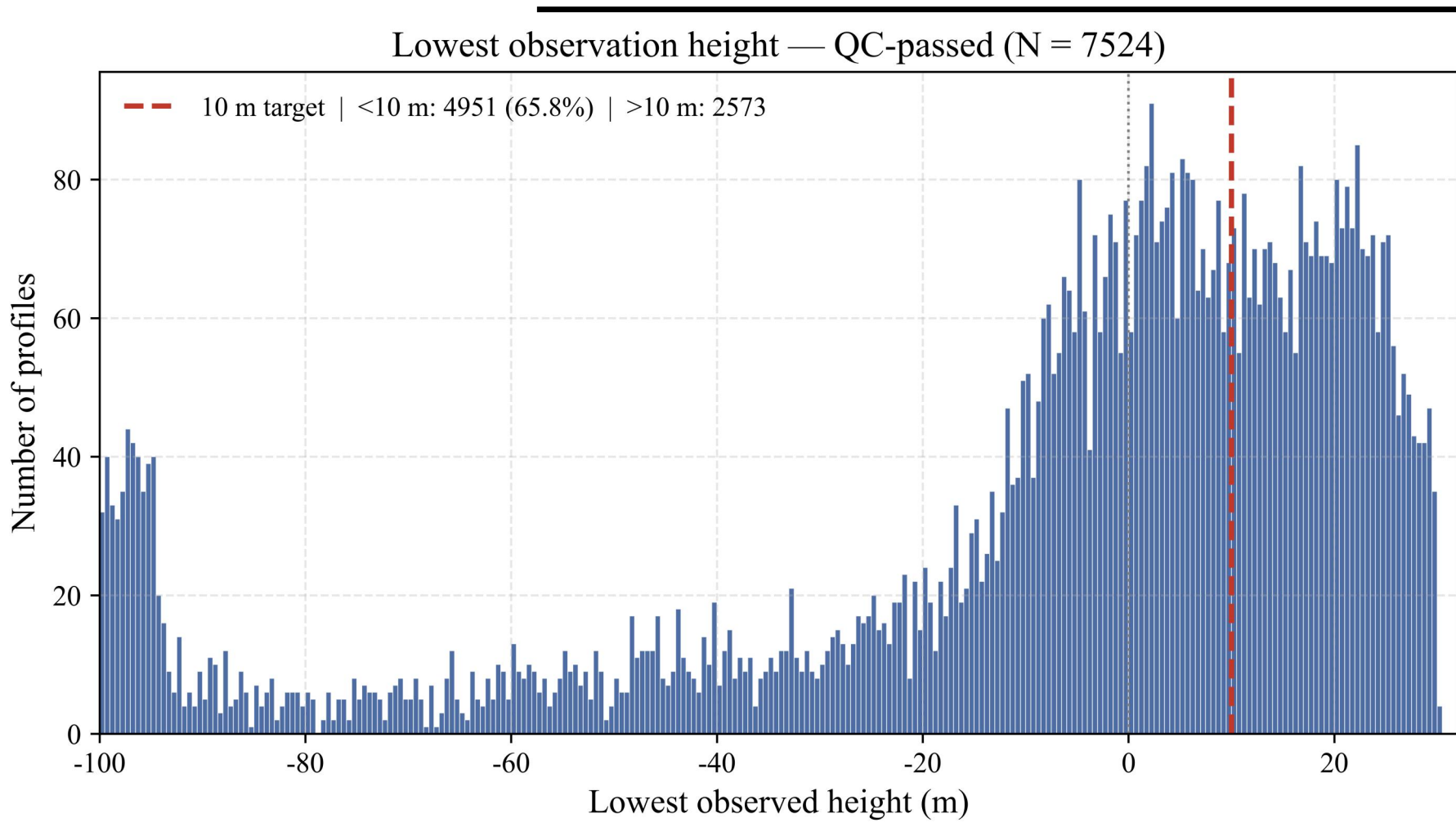

**Figure 1.** Distribution of the lowest observation height across 7,524 QC-passed dropsonde profiles

As shown in Figure 1, 65.8% of profiles (4,951) have their lowest observation below 10 m, for which the 10 m wind is directly sampled. The remaining 34.2% (2,573) have their lowest observation above 10 m but within 30 m; for these, the 10 m wind is assigned from the nearest available observation level. No profile has a lowest observation exceeding 30 m, as the quality-control procedure discards any profile lacking an observation point within ±20 m of 10 m. Profiles with a lowest recorded height below 0 m (i.e., slightly negative values arising from altimeter calibration offsets or sea-surface pressure-height conversion) have the sub-surface level removed prior to analysis; only the valid above-surface portion of each profile is retained. The nearest-neighbor assignment is adopted because the height discrepancy is small, the expected wind-speed difference across this interval is within instrumental uncertainty, and the approach avoids introducing parametric assumptions about friction velocity or roughness length that would be required by any extrapolation scheme.

Additional quality control was applied to remove physically implausible and locally inconsistent wind measurements. Wind samples with either zonal or meridional wind components exceeding ±100 m/s were discarded; this threshold is set above the maximum credible tropical-cyclone wind speed to retain extreme but physically possible observations while excluding obvious sensor errors. Remaining profiles were further screened using a Hampel-type robust outlier filter (Hampel, 1974) applied separately to the zonal and meridional wind components, followed by a robust vertical-gradient consistency check. A record was retained only if the quality-controlled profile contained at least three valid points in each of the 10–100 m, 100–200 m, and 200–300 m layers, and at least nine valid points in total.

After quality control, 7,524 valid profiles from 93 distinct tropical cyclones are retained. The analysis focuses on continuous wind-profile generation between 10 m and 300 m (Ma, 2026).

Figure 2 shows the mean number of dropsonde data points per profile as a function of height (binned in 10 m intervals), stratified by tropical cyclone intensity category (TD, TS, Cat1, Cat2, Cat3, Cat4+). The mean number exceeds 1.0 at all height levels for all intensity categories, indicating that the dropsonde observations provide adequate vertical sampling across the 10–300 m layer. The sampling density decreases gradually with height for all categories, reflecting the typical dropsonde vertical sampling pattern and the QC requirement of at least three valid points per 100 m layer. The relative proportions across categories reflect the underlying distribution of storm intensities in the HRD archive.

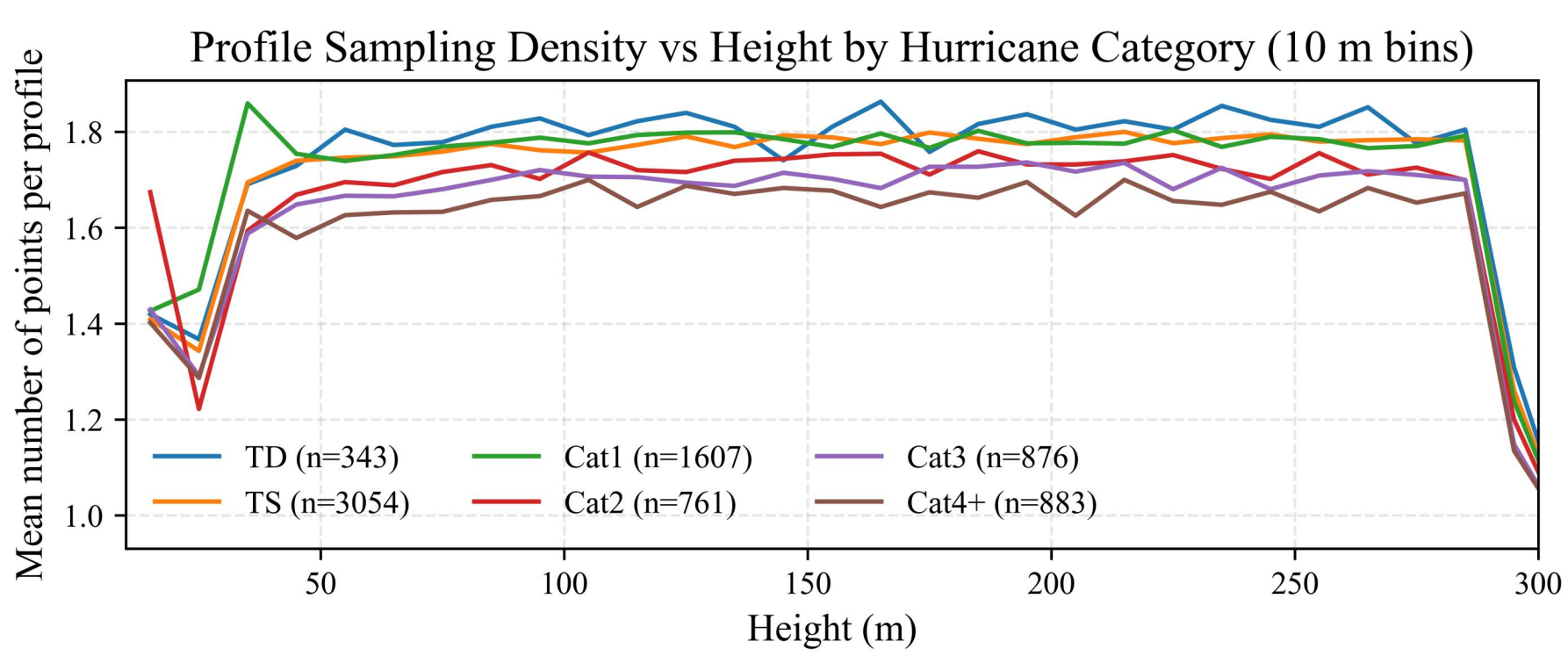

**Figure 2.** Profile sampling density versus height by tropical cyclone intensity category

Figure 3 shows the storm-relative spatial distribution of the 7,524 QC-passed dropsonde profiles. Each point represents the location of one profile in earth-relative azimuth coordinates, with the radial distance normalized by the radius of maximum wind (RMW). In the outer region, the sample density is substantially lower, and model predictions in these sparsely sampled regions carry greater uncertainty. This spatial coverage reflects the typical flight patterns of NOAA hurricane reconnaissance aircraft, which prioritize the inner-core and near-environment regions. The implications of the non-uniform spatial sampling for model confidence are discussed further in Section 5. The large temporal span and broad sampling across storm intensities provide a diverse training dataset for developing a generalized profile reconstruction model.

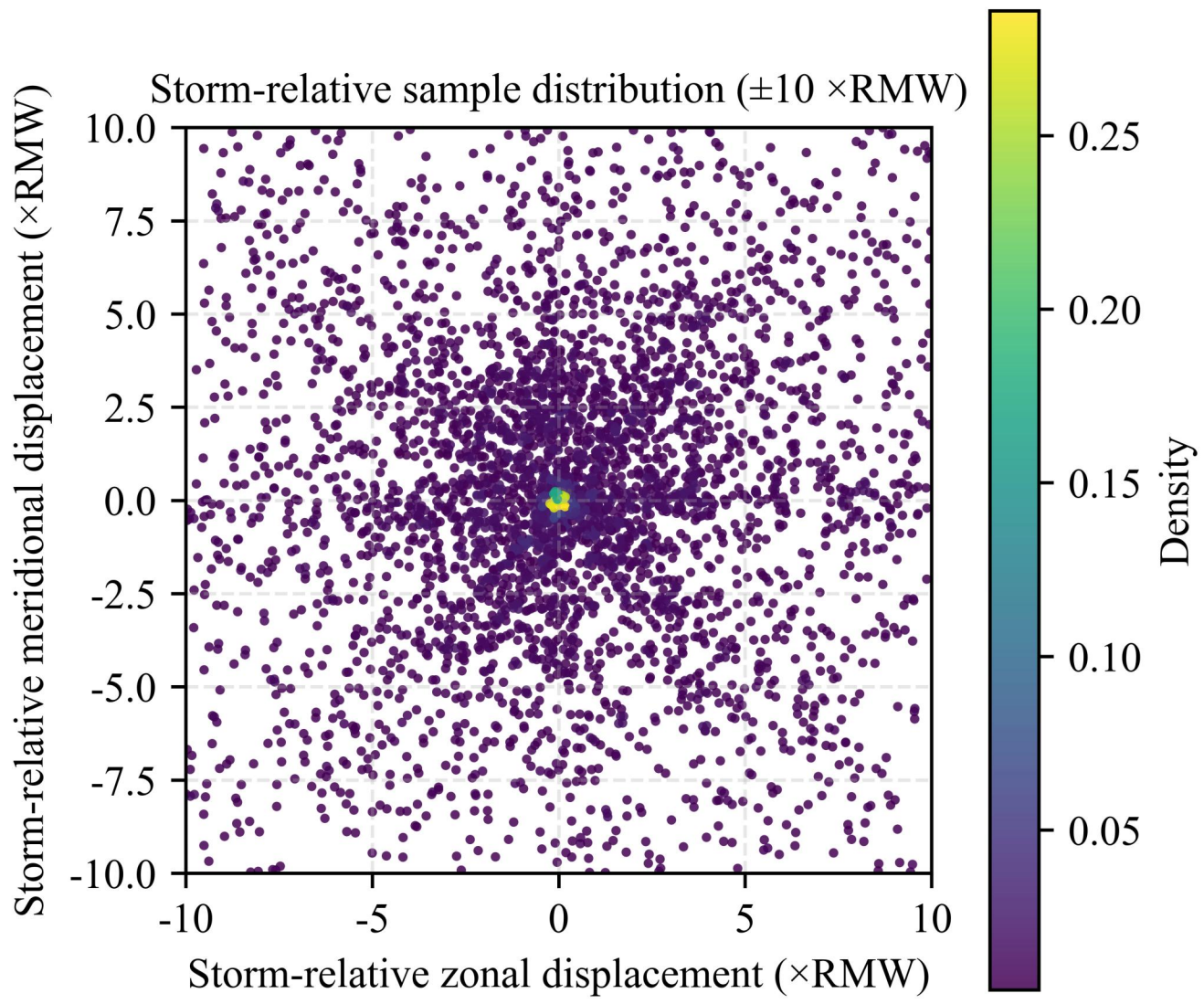

**Figure 3.** Relative position distribution of HRD dropsonde samples in storm-centered coordinates

## 2.2 Coordinate Transformation and Conditional Features

Given the cyclone-center longitude and latitude, denoted by $tclon$ and $tclat$ in the raw records, and the observation-point longitude and latitude, denoted by $lon$ and $lat$, all angular quantities are first converted from degrees to radians. The converted cyclone-center longitude and latitude are denoted by $\lambda_c$ and $\varphi_c$, respectively, and the converted observation-point longitude and latitude are denoted by $\lambda_o$ and $\varphi_o$, respectively. Under a storm-centered local Cartesian approximation, the relative displacements of the observation point with respect to the cyclone center are calculated as:

$$\Delta x = R\cos\phi_c\ (\lambda_o - \lambda_c) \tag{1}$$

$$\Delta y = R\,(\phi_o - \phi_c) \tag{2}$$

where $\Delta x$ is the east-west displacement of the observation point relative to the cyclone center, $\Delta y$ is the north-south displacement of the observation point relative to the cyclone center, and $R$ is the radius of the Earth. The radial distance $r$ and earth-relative azimuth angle (measured counterclockwise from due east) $\alpha$ are then calculated as:

$$r = \sqrt{(\Delta x)^2 + (\Delta y)^2} \tag{3}$$

$$\alpha = \ atan2(\Delta y, \Delta x) \tag{4}$$

The dimensionless radius $\rho$ is obtained by normalizing the radial distance using the radius of maximum wind:

$$\rho = \frac{r}{tcrmw} \tag{5}$$

The sine and cosine of the azimuth angle, $\sin\alpha$ and $\cos\alpha$, are used as model inputs instead of the raw azimuth angle to avoid discontinuities caused by angular periodicity. To obtain the 10 m radial and tangential wind components used as anchoring inputs, the geographic wind components at the reference height are transformed into the storm-centered cylindrical coordinate system. The relationship between the zonal-meridional wind components and the radial-tangential wind components is:

$$U_{10} = V_{r,10}\cos\alpha - V_{t,10}\sin\alpha \tag{6}$$

$$V_{10} = V_{r,10}\sin\alpha + V_{t,10}\cos\alpha \tag{7}$$

The 10 m surface winds are similarly rotated to obtain $\mathrm{V_{r,10}}$ and $\mathrm{V_{t,10}}$. Where $U_{10}$ and $V_{10}$ are the 10 m zonal and meridional wind components, respectively. These transformed wind components are used as the reference-height anchoring variables in the model.

The Coriolis parameter is calculated using the cyclone-center latitude and then decomposed into magnitude and sign:

$$f = 2\Omega\sin(tclat) \tag{8}$$

where Ω is the angular velocity of the Earth, $|f|$ describes the magnitude of rotational influence, and $Sign_f$ represents the hemispheric sign. This decomposition enables the model to sense both absolute rotational strength and the sign difference between the Northern and Southern Hemispheres.

The conditional input features are designed to provide the model with compact but physically meaningful information for inferring vertical profile structure.

In the near-surface layer, the momentum budget is affected by turbulent friction, vertical shear, pressure-gradient forcing, and the Coriolis force. Near the upper part of the boundary layer, a simplified gradient-wind balance can be expressed as:

$$\frac{V_t^2}{r} + fV_t \approx \frac{1}{\rho_a}\frac{\partial p}{\partial r} \tag{9}$$

where $\rho_a$ is the air density, $p$ is pressure, $r$ is radial distance, $V_t$ is tangential wind, and $f$ is the Coriolis parameter. Departures from this approximate balance in the frictional boundary layer induce radial inflow and height-varying inflow angle, motivating the physical constraints used in the training objective.

Based on the above physical considerations, the conditional input features are reconstructed according to four principles:

(1) Scale equivariance: The dimensionless radius $\rho$ is used together with the radius of maximum wind $tcrmw$ to separate the intrinsic storm scale from the storm-relative location.

(2) Rotational equivariance: The sine and cosine of the azimuth angle, $\sin\alpha$ and $\cos\alpha$, are used instead of the raw azimuth angle to avoid discontinuities caused by angular periodicity.

(3) Intensity and environmental modulation: The maximum wind speed $tcwnd$ is used as a proxy for the overall storm intensity, while the cyclone-center latitude and longitude provide large-scale storm-location information.

(4) Reference-height anchoring consistency: The 10 m radial and tangential wind components, $V_{r,10}$ and $V_{t,10}$, are used as reference-height anchoring variables, so that the model output can exactly match the observed 10 m winds through the differential anchoring layer.

The final set of conditional input features is summarized in Table 2. The resulting dataset provides paired environmental predictors and continuous reference wind profiles for training and evaluating the physics-constrained implicit profile network described in the following section.

**Table 2.** Model input features

| Name | Symbol | Unit |
|---|---|---|
| Radial distance (dimensionless) | $\rho$ | — |
| Sine of azimuth | $\sin\alpha$ | — |
| Cosine of azimuth | $\cos\alpha$ | — |
| Maximum wind speed | $tcwnd$ | m/s |
| Radius of maximum wind | $tcrmw$ | m |
| Absolute Coriolis parameter | $\|f\|$ | $s^{-1}$ |
| Sign of Coriolis parameter | $Sign_f$ | — |
| Cyclone center latitude | $tclat$ | degree |
| Cyclone center longitude | $tclon$ | degree |
| 10 m radial wind | $V_{r,10}$ | m/s |
| 10 m tangential wind | $V_{t,10}$ | m/s |

## 3 Model Construction

### 3.1 Problem Formalization

The objective of this study is to reconstruct continuous radial and tangential wind profiles within the tropical cyclone near-surface layer from sparse dropsonde observations and storm-environmental information. Unlike conventional interpolation methods that predict wind only at discrete heights, the proposed framework learns a continuous mapping between storm conditions, height, and wind components, allowing predictions at arbitrary elevations within the observation layer. The target generation height z is normalized to the interval [0, 1] as:

$$\zeta = (\mathrm{z} - 10)/290 \in [0,1] \tag{10}$$

Let $\tilde{\mathbf{x}}$ denote the standardized conditional feature vector listed in Table 2. The network learns the mapping:

$$F_\theta: (\tilde{\mathbf{x}}, \zeta) \mapsto (\hat{V}_r(\zeta), \hat{V}_t(\zeta)) \tag{11}$$

where $\hat{V}_r(\zeta)$ and $\hat{V}_t(\zeta)$ are the reconstructed radial and tangential wind components at height $\zeta$. By evaluating the same conditional vector over a set of target heights, the model generates a continuous near-surface wind profile.

### 3.2 Network Architecture

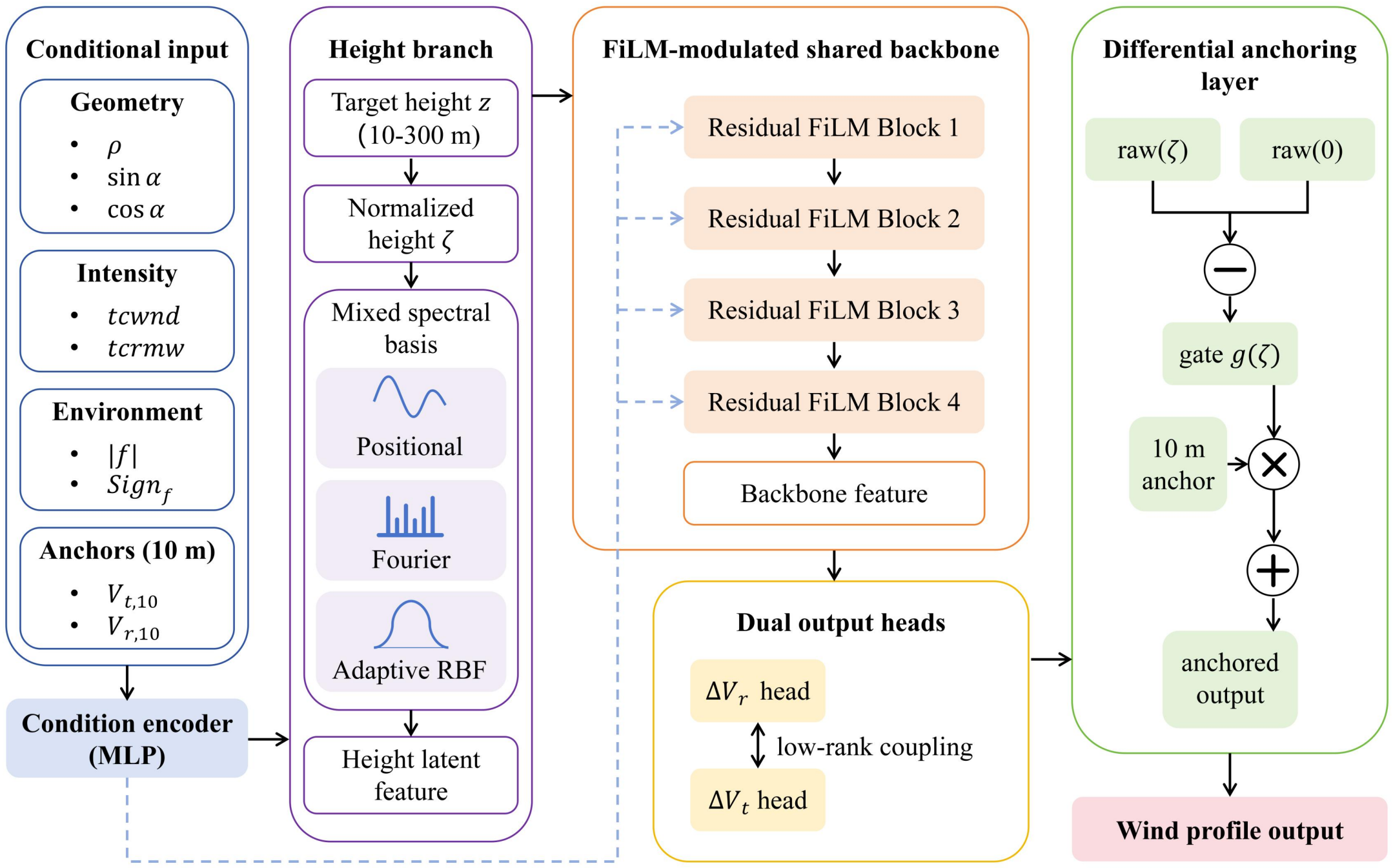

**Figure 4.** Schematic architecture of the Implicit Profile Network (Smartyphoon-BL3D)

As illustrated in Figure 4, The Smartyphoon-BL3D consists of five functional modules: a condition branch, a height branch, a FiLM-modulated shared backbone (Perez et al., 2018), dual output heads, and a differential anchoring layer. This organization separates environmental conditioning, height encoding, cross-layer feature fusion, component-specific output generation, and strict reference-height consistency.

The condition branch maps the standardized conditional feature vector $\tilde{\mathbf{x}}$ into a condition vector $\mathbf{c}$. This vector summarizes storm-relative location, storm scale, intensity, hemispheric rotation, storm-center position, and the 10 m reference winds. The condition vector is then used to generate FiLM modulation parameters and to control the shape and amplitude of the generated profile under different storm environments.

The height branch maps the normalized height $\zeta$ into a high-dimensional latent representation, following the implicit neural representation paradigm in which a continuous function is parameterized by a neural network conditioned on coordinate inputs (Mildenhall et al., 2020; Sitzmann et al., 2020). To capture both smooth vertical trends and localized variations, the height branch uses a mixed spectral basis that combines positional information, Fourier features (Tancik et al., 2020), and radial basis functions:

$$\boldsymbol{\Gamma}_h(\zeta;\rho) = [\zeta, \{\sin(2\pi k\zeta)\}_{k=1}^{K}, \{\cos(2\pi k\zeta)\}_{k=1}^{K}, \{exp(-\frac{(\zeta-c_m)^2}{2\sigma(\rho)^2})\}_{m=1}^{M}] \quad (12)$$

Where $\boldsymbol{\Gamma}_h(\zeta;\rho)$ denotes the height encoding vector, $\zeta$ is the normalized height, $k$ is the Fourier feature index, $m$ is the radial basis function index, $c_m$ is the $m$-th radial basis function center uniformly placed in [0,1], and $\sigma(\rho)$ is the radial basis function width associated with the dimensionless radius $\rho$. The radial basis function width is defined as:

$$\sigma(\rho) = clip(\sigma_0(1 + \alpha_\rho \rho), \sigma_{min}, \sigma_{max}) \quad (13)$$

where $\sigma_0$ is the base width, $\alpha_\rho$ is the adjustment coefficient associated with the dimensionless radius, and $\sigma_{min}$ and $\sigma_{max}$ are the lower and upper bounds of the width, respectively. This design makes the radial basis function width positively correlated with the dimensionless radius. In the inner-core region, where $\rho$ is small, $\sigma(\rho)$ is also small, which helps the model capture sharp near-surface turning. In the outer region, where $\rho$ is larger, $\sigma(\rho)$ increases accordingly, which helps avoid overfitting weak perturbations. The encoded height vector is subsequently projected into the initial height latent representation $\mathbf{h}_0$.

The shared backbone receives the height latent representation and dynamically modulates it using the condition vector. It consists of multiple residual blocks (He et al., 2016) with Feature-wise Linear Modulation (FiLM) (Perez et al., 2018). For the $\ell$-th residual block, the condition vector generates a scaling vector $\boldsymbol{\gamma}_\ell$ and a shifting vector $\boldsymbol{\beta}_\ell$. The scaling vector controls the amplification or suppression of internal features, while the shifting vector changes the feature baseline. The residual update is written as:

$$\mathbf{h}_\ell = \mathbf{h}_{\ell-1} + \mathcal{D}_\ell\left[W_{\ell,2}\psi\left((1 + \boldsymbol{\gamma}_\ell) \odot W_{\ell,1}LN(\mathbf{h}_{\ell-1}) + \boldsymbol{\beta}_\ell\right)\right] \quad (14)$$

where $\mathbf{h}_\ell$ is the output of the$\ell$-th residual block, $\mathbf{h}_{\ell-1}$ is the output of the previous residual block, $W_{\ell,1}$ and $W_{\ell,2}$ are learnable linear mapping matrices, $LN(\cdot)$ denotes layer normalization (Ba et al., 2016), $\psi(\cdot)$ is the activation function, $\boldsymbol{\gamma}_\ell$ is the scaling modulation vector of the $\ell$-th FiLM layer, $\boldsymbol{\beta}_\ell$ is the shifting modulation vector of the $\ell$-th FiLM layer, and $\mathcal{D}_\ell(\cdot)$ denotes stochastic-depth/dropout regularization. The scaling vector $\boldsymbol{\gamma}_\ell$ and shifting vector $\boldsymbol{\beta}_\ell$ are generated from the condition vector $\mathbf{c}$.

To enhance cross-layer information flow, gated skip connections are further introduced in the odd residual blocks. These connections fuse the initial height representation with the current residual-block output:

$$\mathbf{h}_\ell \leftarrow \mathbf{h}_\ell + \theta(\mathbf{g}_\ell) \odot \mathbf{P}_\ell \mathbf{h}_0, \quad \ell \in oddlayers \quad (15)$$

where $\mathbf{g}_\ell$ is the gate parameter vector of the $\ell$-th odd residual block, $\theta(\cdot)$ is the sigmoid function, $\mathbf{P}_\ell$ is a linear projection matrix, and $\mathbf{h}_0$ is the initial output of the height branch. The final output of the shared backbone is denoted as $\mathbf{s}(\zeta)$.

The dual output heads generate the unanchored radial and tangential wind increments from the shared semantic representation $\mathbf{s}(\zeta)$. To introduce necessary cross-component physical dependence while preserving the advantage of component-wise decoupling, a low-rank coupling term is added:

$$\begin{bmatrix} \Delta V_r \\ \Delta V_t \end{bmatrix} = \begin{bmatrix} w_r^\top \\ w_t^\top \end{bmatrix} \mathbf{s}(\zeta) + \mathbf{C}\, g(s(\zeta)), \quad rank(\mathbf{C}) \leq 2 \quad (16)$$

where $w_r$ is the weight vector for the radial wind increment $\Delta V_r$, $w_t$ is the weight vector for the tangential wind increment $\Delta V_t$, $\mathbf{s}(\zeta)$ is the shared semantic representation, $g(\mathbf{s}(\zeta))$ denotes the nonlinear feature term obtained from the shared semantic representation, and $\mathbf{C}$ is the low-rank coupling matrix. The low-rank matrix is regularized with low rank and small magnitude, enabling the model to represent cross-component physical relationships, such as height-varying inflow angle, while retaining the advantages of component-wise decoupling.

The differential anchoring layer converts the radial and tangential wind increments into the final wind components. This layer ensures that the model outputs at the normalized reference height $\zeta = 0$ are exactly equal to the input radial and tangential winds at the reference height. When the reference height is set to 10 m, the final output for $q \in \{r, t\}$ is written as:

$$\widehat{V}_q(\zeta) = V_{q,10} + \Delta V_q(\zeta) - \Delta V_q(0), \quad q \in \{r, t\} \tag{17}$$

where $\widehat{V}_q(\zeta)$ denotes the reconstructed radial or tangential wind component, $V_{q,10}$ denotes the 10 m radial or tangential wind component, $\Delta V_q(\zeta)$ denotes the reconstructed radial or tangential wind increment at normalized height $\zeta$, and $\Delta V_q(0)$ denotes the corresponding increment at $\zeta = 0$. This design guarantees that $\widehat{V}_r(0) = V_{r,10}$ and $\widehat{V}_t(0) = V_{t,10}$ hold exactly.

### 3.3 Loss Function and Physical Regularization

The training objective is designed as a weighted combination of data-fitting and physically motivated regularization terms. These terms jointly improve data fidelity, suppress non-physical oscillations, enforce hemispheric consistency, guide inflow-angle behavior, introduce near-surface logarithmic-law information, and stabilize the modulation and coupling components of the network. This design follows the physics-informed neural network paradigm, in which physical knowledge is embedded into the training objective as auxiliary regularization terms (Raissi et al., 2019). The total loss function $L$ is defined as the weighted sum of the following terms:

$$L = \lambda_{fit}L_{fit} + \lambda_{curv}L_{curv} + \lambda_{hemi}L_{hemi} + \lambda_{\varphi}L_{angle} + \lambda_{log}L_{log} + \lambda_{film}L_{film} + \lambda_{int}L_{int} + \lambda_{d}L_{slope} \tag{18}$$

The data-fitting loss adopts a weighted Huber penalty (Huber, 1964) to reduce sensitivity to outliers and to assign larger weights to near-surface and high-curvature regions

$$L_{fit} = \frac{1}{\sum_{j=1}^{J} w_j} \sum_{j=1}^{J} w_j \left[ \rho_\delta \left( \widehat{V}_r(z_j) - V_r(z_j) \right) + \rho_\delta \left( \widehat{V}_t(z_j) - V_t(z_j) \right) \right] \tag{19}$$

where $J$ is the total number of sampled height points, $z_j$ is the height of the $j$-th sampled point, $w_j$ is the sample weight, $\rho_\delta(\cdot)$ is the Huber penalty function, $\widehat{V}_r(z_j)$ and $V_r(z_j)$ are the reconstructed and observed radial wind components at height $z_j$, and $\widehat{V}_t(z_j)$ and $V_t(z_j)$ are the reconstructed and observed tangential wind components at height $z_j$, respectively.

To suppress non-physical high-frequency oscillations, a height-dependent curvature regularization term is introduced:

$$L_{\text{curv}} = \frac{1}{J-2} \sum_{j=2}^{J-1} w_{\text{curv}}(z_j) \left[ \rho_{\delta_c}\left( \Delta^2 \widehat{V}_r(z_j) \right) + \rho_{\delta_c}\left( \Delta^2 \widehat{V}_t(z_j) \right) \right] \tag{20}$$

where $w_{\text{curv}}(z_j)$ is the curvature weight at height $z_j$, $\Delta^2$ denotes the second-order finite-difference operator along the height direction, and $\rho_{\delta_c}(\cdot)$ is the robust penalty function with threshold parameter $\delta_c$.

Based on hemispheric circulation characteristics, a tangential-wind sign consistency constraint is imposed:

$$L_{\text{hemi}} = \frac{1}{J}\sum_{j=1}^{J} \max\left(0, -\, sign_f \widehat{V}_t(z_j)\right) \tag{21}$$

where $sign_f$ is the sign of the Coriolis parameter.

An inflow-angle prior is used to constrain the adjacent-height variation of the inflow angle in the boundary layer:

$$L_{\text{angle}} = \frac{1}{J'}\sum_{z_j \in [10,300]\text{m}} \max\left(0, -\, \Delta\varphi(z_j)\right) \tag{22}$$

where $J'$ is the number of valid sampled points participating in the inflow-angle prior loss within the 10-300 m interval, and $\Delta\varphi(z_j)$ denotes the inflow-angle variation between adjacent height levels.

Because the near-surface wind-speed magnitude approximately follows the logarithmic law, a soft prior is introduced within the 10-40 m layer:

$$L_{\log} = \frac{1}{J_{\text{ns}}}\sum_{z_j \in [10,40]\text{m}} \rho_{\delta_l}\left(\left\|\widehat{\mathbf{V}}(z_j)\right\|_2 - U_{\log}(z_j)\right) \tag{23}$$

$$U_{\log}(z) = \frac{u_*}{\kappa}\ln\left(\frac{z}{z_0}\right) \tag{24}$$

where $\widehat{\mathbf{V}}(z_j) = [\widehat{V}_r(z_j), \widehat{V}_t(z_j)]^T$ is the reconstructed wind vector at height $z_j$, $J_{\text{ns}}$ is the number of valid sampled points participating in the logarithmic-law soft prior within the 10-40 m interval, $\rho_{\delta_l}(\cdot)$ is the robust penalty function with threshold parameter $\delta_l$, $U_{\log}(z_j)$ is the reference wind-speed magnitude derived from the near-surface logarithmic law, $u_*$ is the friction velocity, $\kappa$ is the Kármán constant, and $z_0$ is the surface roughness length.

The logarithmic-law prior is applied only in the 10–40 m layer because this interval is closest to the 10 m anchoring height and represents the lowest near-surface part of the analyzed profile, where surface-layer scaling is most relevant. The upper bound of 40 m is not treated as a sharp physical interface; rather, it is used as a conservative height limit to provide weak near-surface guidance without imposing a logarithmic shape on the higher tropical-cyclone boundary layer. Above this layer, curvature, height-dependent inflow turning, low-level jets, and storm-asymmetric structures become more pronounced, so the profile shape is learned mainly from observations and other physical regularization terms.

To stabilize the conditional modulation process, a regularization term is imposed on the FiLM scaling parameters:

$$L_{\text{film}} = \frac{1}{N_{\text{FiLM}}}\sum_{\ell=1}^{N_{\text{FiLM}}} \|\boldsymbol{\gamma}_\ell\|_2^2 \tag{25}$$

where $N_{\text{FiLM}}$ is the total number of FiLM modulation layers, $\ell$ indexes the FiLM layer, and $\boldsymbol{\gamma}_\ell$ is the scaling modulation vector of the $\ell$-th FiLM layer.

To suppress unnecessary cross-component coupling, the low-rank coupling matrix is regularized as:

$$L_{\text{int}} = \|\mathbf{C}\|_F^2 \qquad (26)$$

where $\mathbf{C}$ is the low-rank coupling matrix and $\|\cdot\|_F$ denotes the Frobenius norm.

Finally, a slope regularization term is introduced to control excessive first-order vertical variation:

$$L_{\text{slope}} = \frac{1}{J-1}\sum_{j=1}^{J-1}\left\|\frac{\widehat{\mathbf{V}}(z_{j+1}) - \widehat{\mathbf{V}}(z_j)}{z_{j+1} - z_j}\right\|_2^2 \qquad (27)$$

where $\widehat{\mathbf{V}}(z_j)$ is the reconstructed wind vector at height $z_j$.

## 4 Results and Analysis

### 4.1 Experimental Setup

All experiments were conducted on a computer equipped with an NVIDIA GeForce RTX 5070 GPU with 8 GB memory. The dataset was divided into training, validation, and test sets at a ratio of 8:1:1 using a cyclone-wise grouped and intensity-stratified sampling strategy. Samples from the same tropical cyclone were assigned to only one subset, thereby preventing information leakage caused by placing profiles from the same storm in both training and testing. The training set was used for parameter learning, the validation set was used for hyperparameter selection and early stopping, and the test set was reserved exclusively for final performance reporting. The model was trained and evaluated on wind-profile data spanning heights from 10 m to 300 m. All feature-normalization statistics were computed using only the training set and then fixed during validation and testing.

In summary, the trained Smartyphoon-BL3D takes as input the storm intensity (maximum wind speed), storm-center latitude and longitude, the storm-relative radial distance (normalized by RMW), the earth-relative azimuth angle, and the observed 10-m radial and tangential winds, and outputs the reconstructed radial and tangential wind components at any target height between 10 m and 300 m at the specified storm-relative location. The model is thus a continuous conditional mapping from compact storm-environmental information to vertical wind profiles within approximately 10 × RMW of the storm center.

In this section, we systematically evaluate the performance of the proposed model on the independent test set from four perspectives: overall parity and correlation performance, inflow-angle error distribution, representative vertical-profile morphology, and intensity-stratified error characteristics. The evaluation variables include radial wind, tangential wind, wind-speed magnitude, and wind direction or inflow angle. For intensity-stratified statistics, mean absolute error (MAE), mean error (ME), root mean square error (RMSE), percentage bias (Bias%), and coefficient of determination ($R^2$) are reported.

### 4.2 Overall Reconstruction Consistency and Correlation Performance

As shown in Figure 5, the model outputs exhibit strong linear agreement with observations across the key variables. The scatter plots for radial wind, tangential wind, and wind-speed magnitude are tightly clustered around the 1:1 reference line, with coefficients of determination reaching $R^2 = 0.948$ for radial wind, $R^2 = 0.963$ for tangential wind, and $R^2 = 0.963$ for wind-speed magnitude. These results demonstrate that the model effectively captures the dominant variability and magnitude characteristics of the boundary-layer wind field. The wind direction or inflow angle also shows good agreement over the main observational range, indicating that the model learns both component magnitude and directional structure.

The agreement in both radial and tangential components suggests that the conditional modulation and low-rank coupling modules help preserve the relationship between secondary radial circulation and primary tangential circulation. Together with the differential anchoring layer, this design provides a stable basis for continuous wind-profile generation and subsequent stratified error analysis.

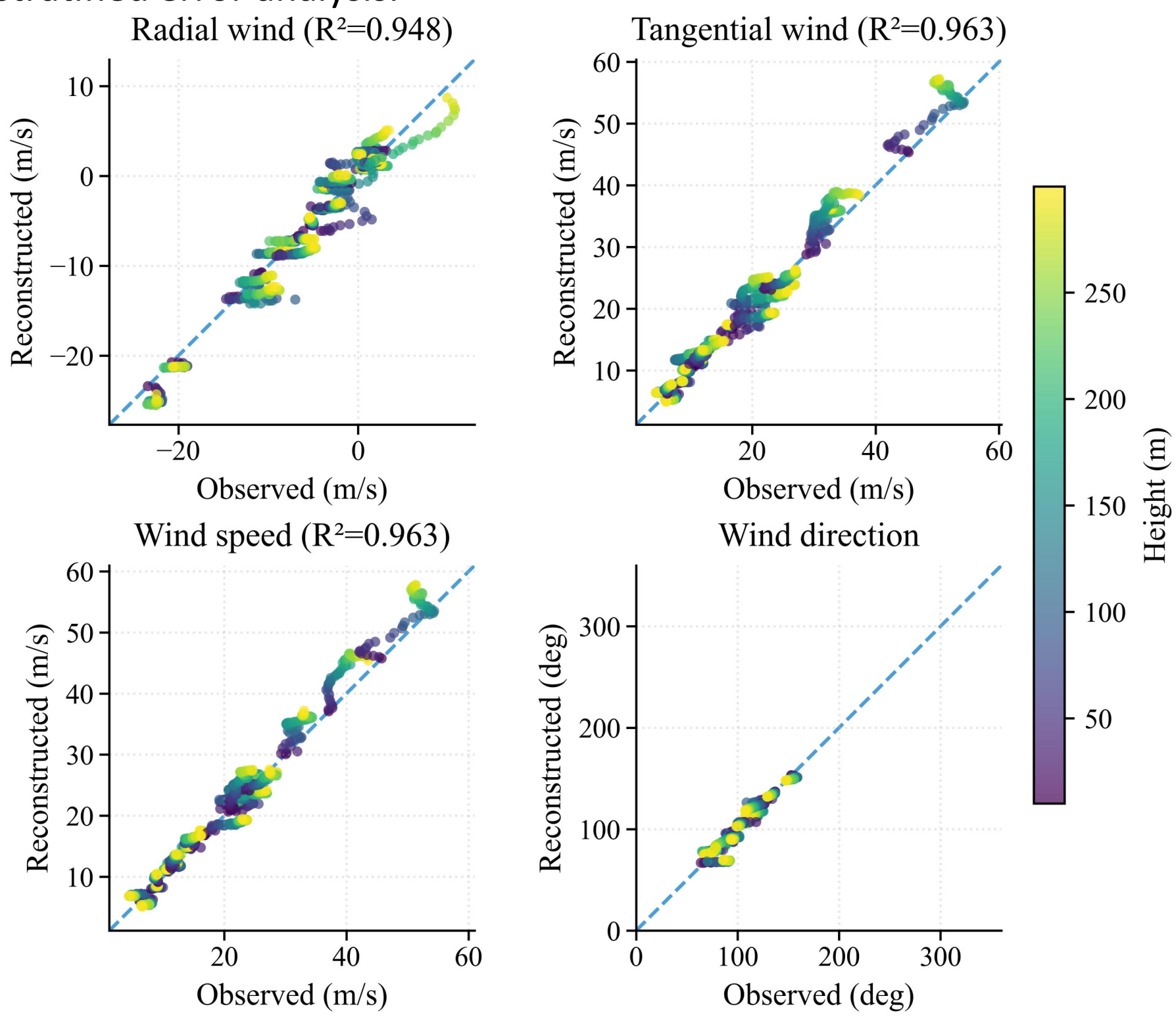

**Figure 5.** Overall parity plots for radial wind, tangential wind, wind-speed magnitude, and wind direction

### 4.3 Inflow Angle Error Distribution Characteristics

As shown in Figure 6, the inflow-angle parity plot and error histogram further reveal the structure of directional reconstruction errors. Overall, the reconstructed inflow angles maintain high consistency with observations, and the error distribution is approximately unimodal and centered near 0°. This indicates that most directional errors are small random deviations rather than large-scale systematic biases. A small number of tail errors remain, suggesting the presence of relatively large angular deviations in limited samples. These tail errors may be associated with two factors. First, under strong convection, rapid structural evolution, or enhanced local shear, the inflow angle may vary sharply with height and azimuth, increasing the difficulty of model learning and generalization. Second, fine-scale boundary-layer structures such as boundary layer rolls (Morrison et al., 2005; Zhang et al., 2008) may contribute to localized inflow-angle deviations that are not fully captured by the current conditioning variables. Prior studies have shown that inflow angle varies systematically with storm motion and environmental vertical wind shear (Zhang & Uhlhorn, 2012; Uhlhorn et al., 2014; Ming et al., 2022), suggesting that incorporating these factors as additional conditioning variables may reduce the residual directional errors.

Because the inflow angle is derived from the reconstructed radial and tangential winds rather than directly output as an independent variable, the overall agreement also suggests that the two component heads remain mutually consistent. This is important for applications in which wind direction, surface stress, and momentum flux are computed from the generated wind vector.

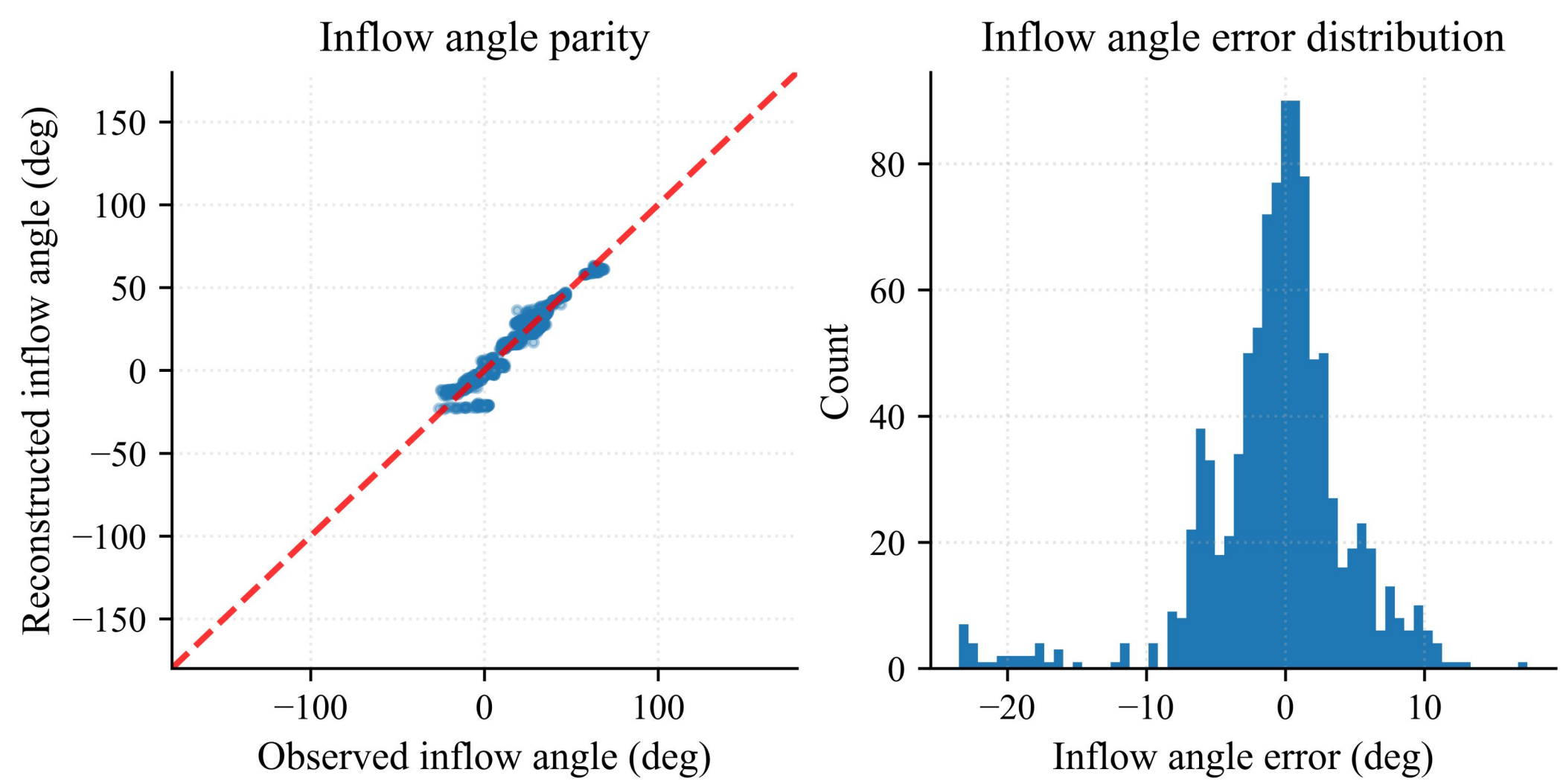

**Figure 6.** Inflow-angle parity and error-distribution characteristics

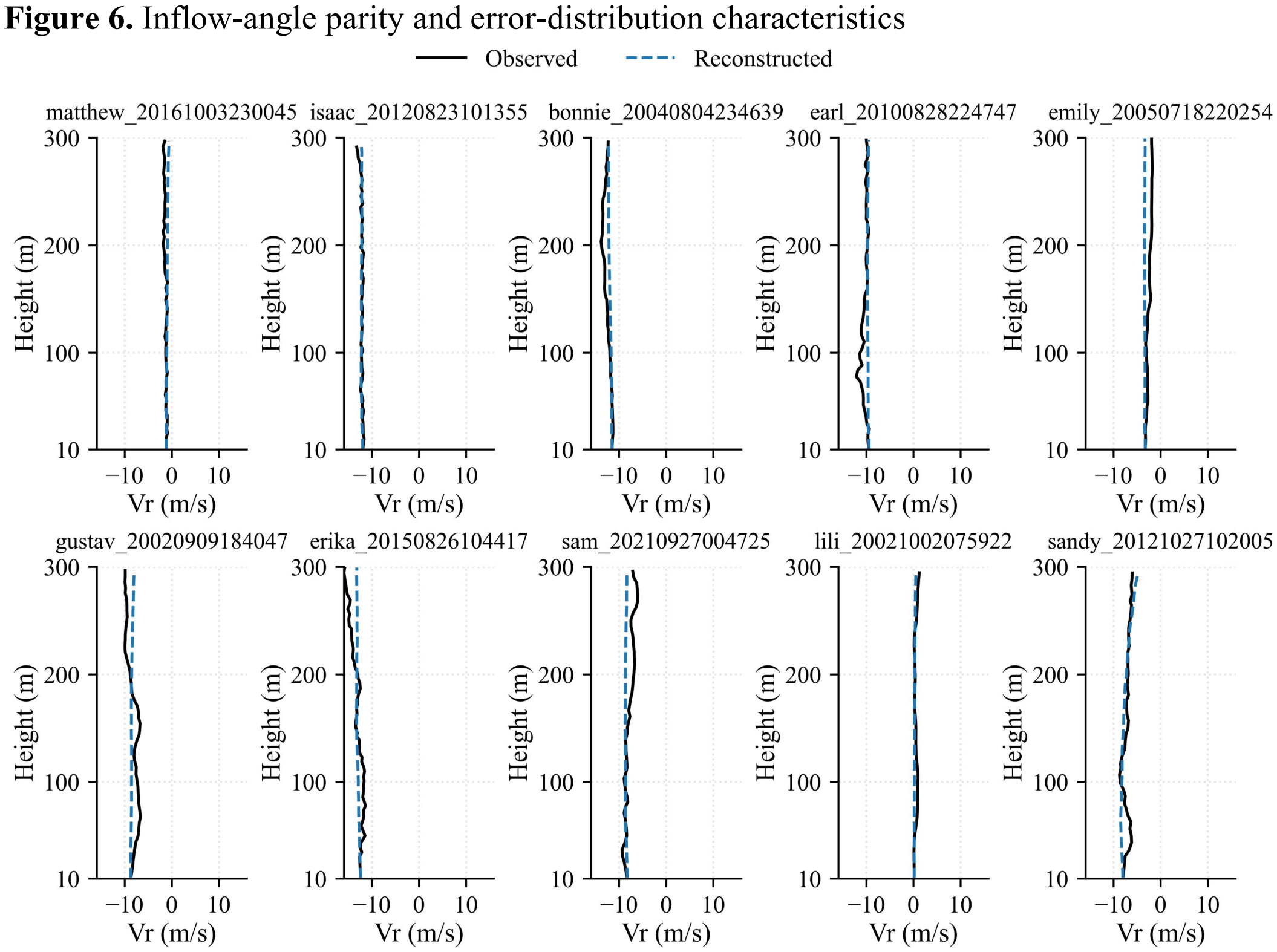

**Figure 7.** Representative profile comparisons for radial wind

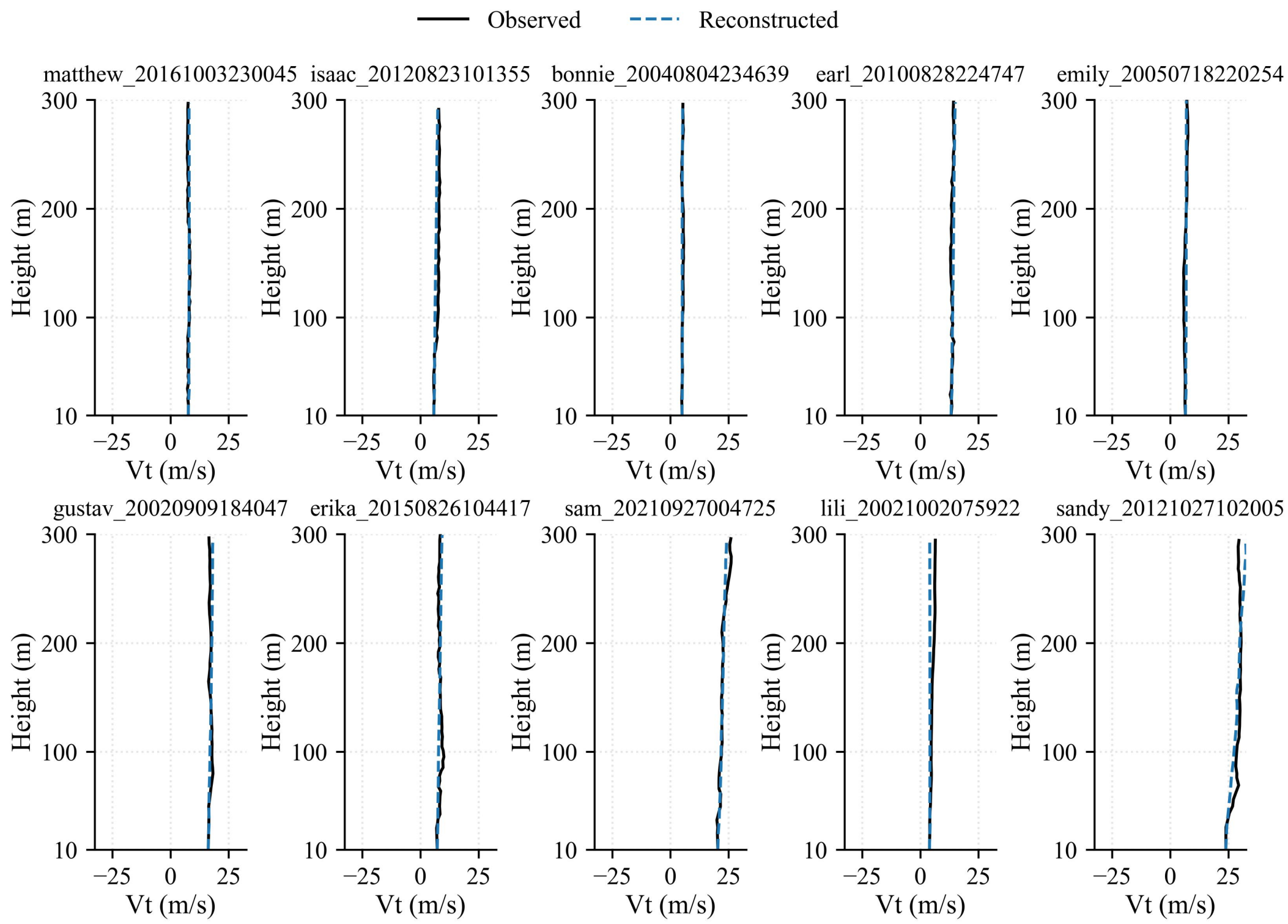

**Figure 8.** Representative profile comparisons for tangential wind

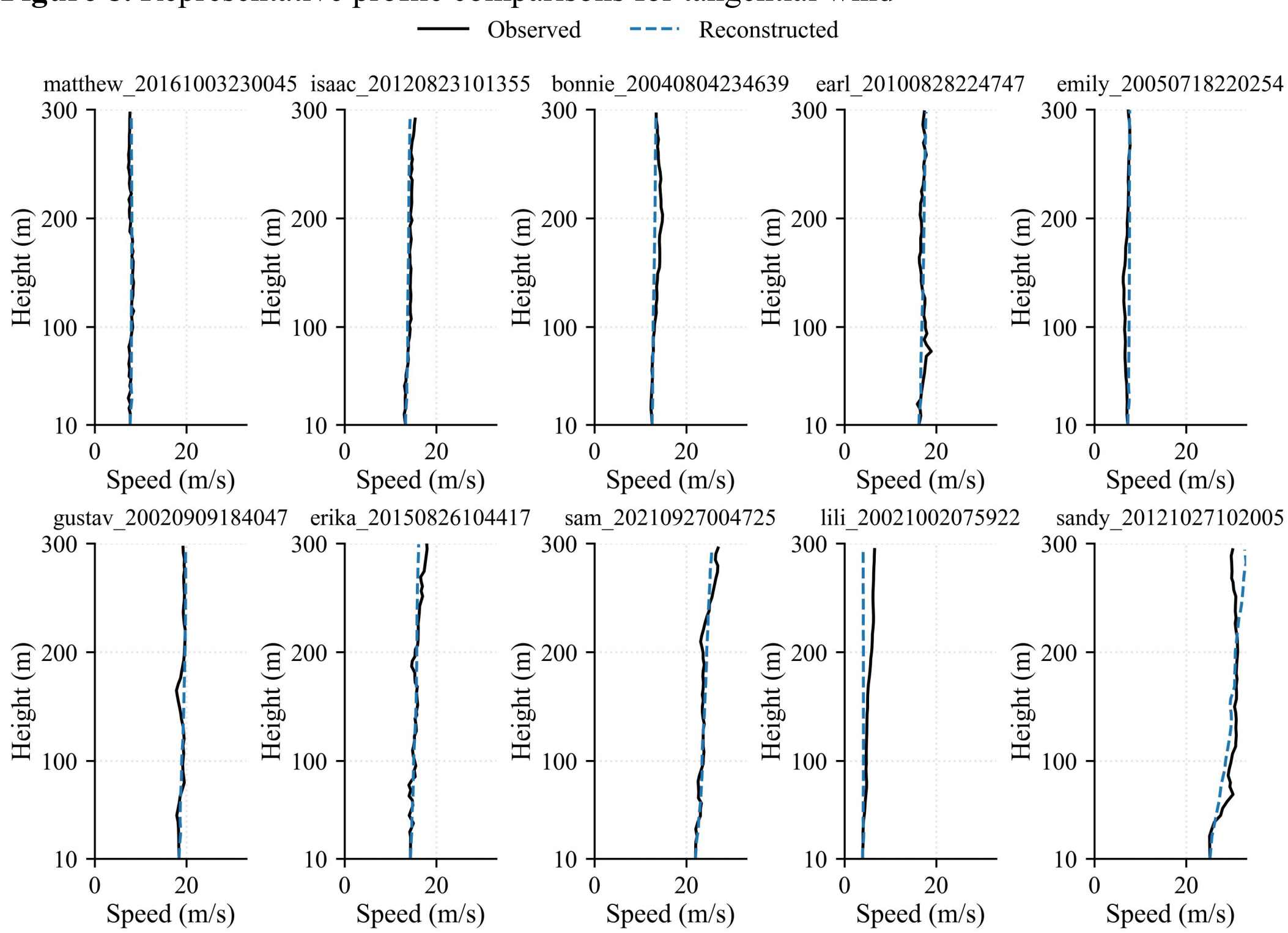

**Figure 9.** Representative profile comparisons for wind-speed magnitude

## 4.4 Representative Profile Comparisons

To examine whether the model reproduces vertical morphology rather than only point-wise statistics, ten individual profiles were randomly selected from the independent test set for observation-reconstruction comparisons. These cases were not chosen a posteriori according to reconstruction accuracy and are therefore not intended to represent a specially favorable or extreme event. Instead, they provide ordinary sample-level examples for assessing whether the Smartyphoon-BL3D can reproduce typical vertical profile shapes under realistic test conditions. The storm name and observation time of each selected profile are shown in the title of each subplot. Figures 7–9 show the corresponding comparisons for radial wind, tangential wind, and wind-speed magnitude. The reconstructed curves largely coincide with observations across most height levels. In the middle and lower layers, the model captures the overall slope, inflection locations, and amplitude variations of the profiles, indicating that the continuous representation preserves the vertical organization of the boundary-layer wind field.

Slight deviations are still observed in some near-surface or upper-level portions of individual cases. These deviations may be associated with local turbulence, strong shear layers, rapidly evolving storm asymmetries, or insufficient representation of rare extreme structures. Nevertheless, the overall curve-level consistency indicates that the mixed spectral height branch and FiLM-modulated shared backbone describe continuous vertical variation rather than merely fitting isolated height points.

Furthermore, as shown in Figure 10, the comparison of mean profiles and interquartile ranges (IQR, 25th-75th percentiles) shows that the reconstructed mean curves are consistent with the observed mean curves in terms of the overall vertical trend. The reconstructed dispersion range also overlaps substantially with the observed IQR, indicating that the model captures not only the central tendency but also the typical variability of the samples. Height ranges with wider observed IQRs often correspond to more complex dynamical structures, such as low-level jets, strong shear layers, or local wind maxima, where reconstructive uncertainty is also larger. This suggests that future improvements may benefit from uncertainty modeling and enhanced learning of extreme or sparsely sampled cases.

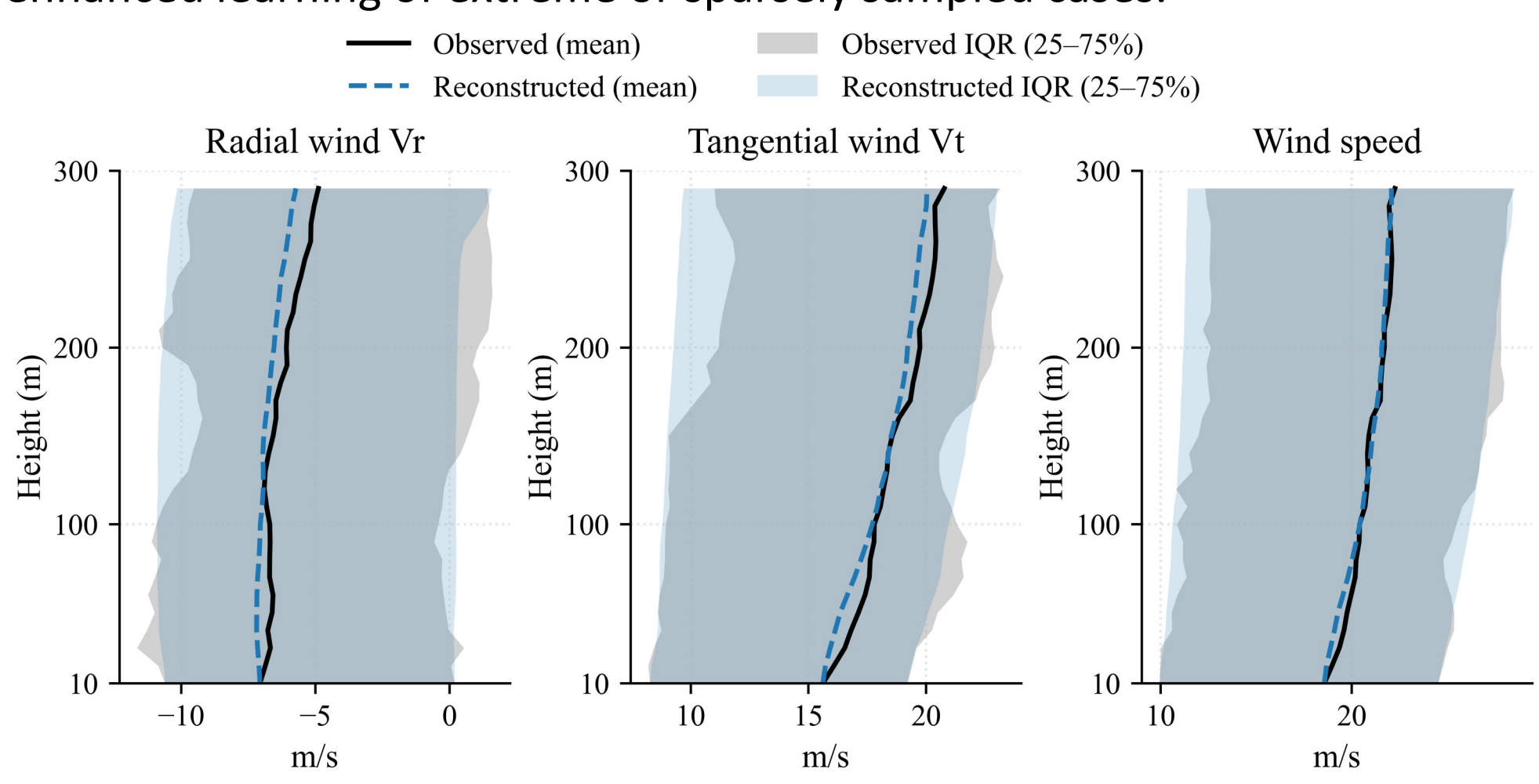

**Figure 10.** Mean profile and interquartile range (IQR) comparison

## 4.5 Error Characteristics by Intensity Category

To quantitatively assess the model's applicability across different storm intensities, we computed error metrics, including MAE, ME, RMSE, Bias%, and $R^2$, for both radial wind and tangential wind stratified by intensity category, including TD, TS, Cat1, Cat2, Cat3, and Cat4+. These statistics are further analyzed together with the height-dependent RMSE profiles to diagnose how model errors vary with storm intensity and height.

As shown in Table 3, the reconstruction errors for radial wind generally increase with storm intensity. The MAE rises from 0.722 m/s for TD to 1.961 m/s for Cat4+, while the RMSE increases from 0.975 m/s to 3.277 m/s. Although the increase is not strictly monotonic, as the Cat3 error is slightly lower than that of Cat2, the overall pattern indicates that stronger storms are more difficult to reconstruct. This is likely related to the stronger radial inflow/outflow structures (Fischer et al., 2022; Zhang et al., 2022) and greater spatial heterogeneity associated with intense tropical cyclones. We note that the larger absolute errors for stronger storms partly reflect the larger background wind speeds; reporting normalized error metrics (e.g., RMSE normalized by the category-mean wind speed) alongside absolute errors may provide complementary insight and is recommended for future work. Despite the larger errors at higher intensities, the $R^2$ values remain high across all categories, ranging from 0.938 to 0.966, indicating that the model captures the dominant variability of radial wind well. The bias values are generally small. TD and TS show slight underestimation, with ME values of -0.105 m/s and -0.190 m/s, whereas the hurricane categories show weak positive bias. For Cat4+, the ME is 0.187 m/s and the Bias% is 1.47%, suggesting a slight overestimation rather than underestimation at the highest intensity level.

As shown in Table 4, tangential wind also exhibits an overall increase in reconstruction error with storm intensity. The MAE increases from 0.786 m/s for TD to 1.924 m/s for Cat4+, and the RMSE increases from 0.976 m/s to 3.193 m/s. Similar to radial wind, the trend is not perfectly monotonic, since the Cat3 errors are slightly lower than those of Cat2. Nevertheless, the larger errors for stronger storms suggest that intense tangential circulation, sharp inner-core wind maxima, and steep vertical gradients increase the difficulty of reconstruction. The $R^2$ values remain consistently high, ranging from 0.951 to 0.967, and reach 0.964 for Cat4+, demonstrating that the model maintains strong capability in representing the dominant tangential wind structure across all intensity categories. The bias statistics indicate a weak underestimation tendency from TD to Cat3, with negative ME values, while Cat4+ shows a small positive bias, with ME = 0.264 m/s and Bias% = 1.10%. Overall, the bias magnitudes are limited, suggesting that the reconstruction errors are mainly associated with increased variability and extreme wind structures rather than strong systematic bias.

**Table 3.** Reconstruction error statistics for radial wind grouped by intensity category

| Category | MAE (m/s) | ME (m/s) | RMSE (m/s) | Bias% | $R^2$ |
|---|---|---|---|---|---|
| TD | 0.722 | -0.105 | 0.975 | -2.38 | 0.966 |
| TS | 1.084 | -0.190 | 1.559 | -3.01 | 0.951 |
| Cat1 | 1.379 | 0.175 | 2.099 | 2.23 | 0.941 |
| Cat2 | 1.649 | 0.065 | 2.397 | 0.64 | 0.967 |
| Cat3 | 1.526 | 0.175 | 2.341 | 1.84 | 0.946 |
| Cat4+ | 1.961 | 0.187 | 3.277 | 1.47 | 0.938 |

**Table 4.** Reconstruction error statistics for tangential wind grouped by intensity category

| Category | MAE (m/s) | ME (m/s) | RMSE (m/s) | Bias% | $R^2$ |
|---|---|---|---|---|---|
| TD | 0.786 | -0.113 | 0.976 | -2.19 | 0.957 |
| TS | 1.154 | -0.167 | 1.691 | -1.48 | 0.951 |
| Cat1 | 1.443 | -0.044 | 2.042 | -0.26 | 0.961 |
| Cat2 | 1.823 | -0.031 | 2.687 | -0.15 | 0.967 |
| Cat3 | 1.649 | -0.301 | 2.541 | -1.50 | 0.962 |
| Cat4+ | 1.924 | 0.264 | 3.193 | 1.10 | 0.964 |

As illustrated in Figures 11–13, the height-dependent RMSE profiles show that reconstruction errors generally increase with storm intensity, but the relationship is not strictly monotonic. Weak systems such as TD and TS maintain relatively low RMSE throughout the 10–300 m layer, whereas stronger categories usually show larger errors and stronger height dependence. The Cat4+ category consistently produces the largest RMSE values in all three variables, indicating that the most intense storms are the most difficult to reconstruct.

For radial wind, the RMSE increases from TD and TS toward stronger categories, with Cat4+ showing the highest errors across most heights. However, Cat3 exhibits a clear non-monotonic behavior: although it is a stronger category than Cat2, its RMSE is comparable to or even lower than Cat2 at several height levels. This suggests that radial-wind reconstruction error is not controlled by storm intensity alone, but is also influenced by sample distribution, storm-relative location, and the complexity of the sampled boundary-layer structure.

For tangential wind, the same overall intensity dependence is observed. TD and TS show relatively small errors, while Cat2 and Cat4+ have larger RMSE values. Cat4+ remains clearly separated from the other categories, especially in the upper part of the 10–300 m layer. Cat3 again does not follow a strictly intensity-ordered pattern, with RMSE values close to or lower than Cat2 at some heights. This may indicate that the Cat3 cases contain smoother tangential-wind profiles or fewer extreme inner-core wind maxima than the Cat4+ samples.

For wind-speed magnitude, the RMSE profiles also increase overall with storm intensity, and Cat4+ shows the largest errors with a clear increase at higher levels. Nevertheless, Cat3 remains lower than Cat4+ and is not consistently higher than Cat2, further confirming the non-monotonic behavior among strong categories. Therefore, the increased error in intense storms is physically interpretable, but category-based errors should be interpreted together with sample representativeness, storm-relative sampling location, eyewall structure, and vertical shear. Reconstruction errors increase systematically with storm intensity, particularly for major hurricanes, reflecting the increasingly complex and rapidly varying boundary-layer structure associated with stronger storms.

Overall, Figures 11–13 indicate that the model captures the main wind-profile structure across intensity categories, but reconstruction becomes more challenging for intense and dynamically complex storms. The consistently high errors for Cat4+ reflect the effects of stronger wind gradients, sharper eyewall structures, enhanced asymmetry, and strong local vertical shear within the 10–300 m layer (i.e., the change in wind speed from the surface to approximately 300 m, which intensifies with storm strength). This is distinct from the environmental vertical wind shear (typically defined over the 850–200 hPa deep-tropospheric layer), which modulates large-scale vortex asymmetries but is not explicitly resolved in the current conditioning variables.

Meanwhile, the relatively moderate Cat3 errors highlight that reconstruction uncertainty depends not only on storm intensity, but also on the detailed structure and sampling characteristics within each category. These results indicate that major hurricanes remain the most challenging reconstruction environment because of their highly nonlinear boundary-layer dynamics.

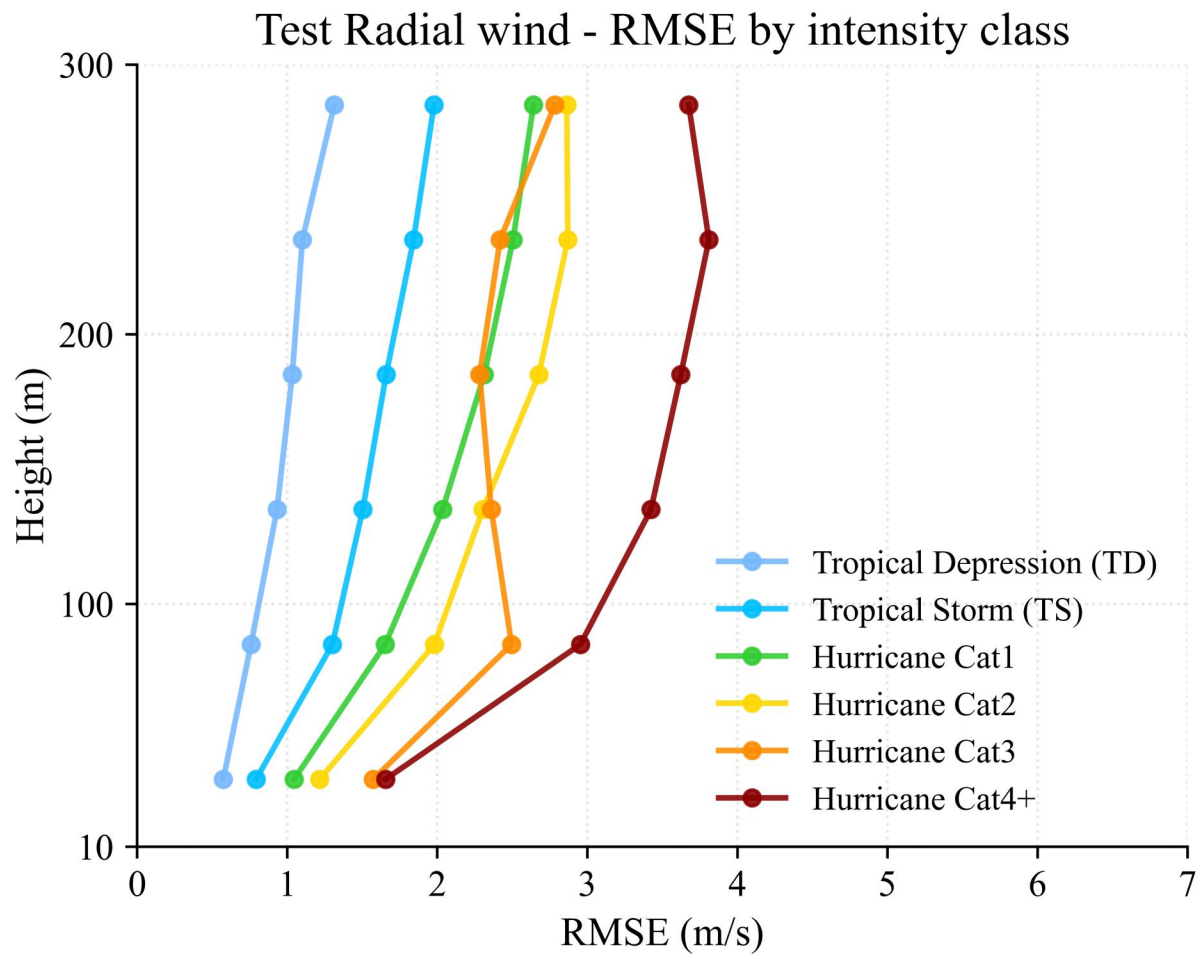

**Figure 11.** Height-dependent RMSE for radial wind by intensity category

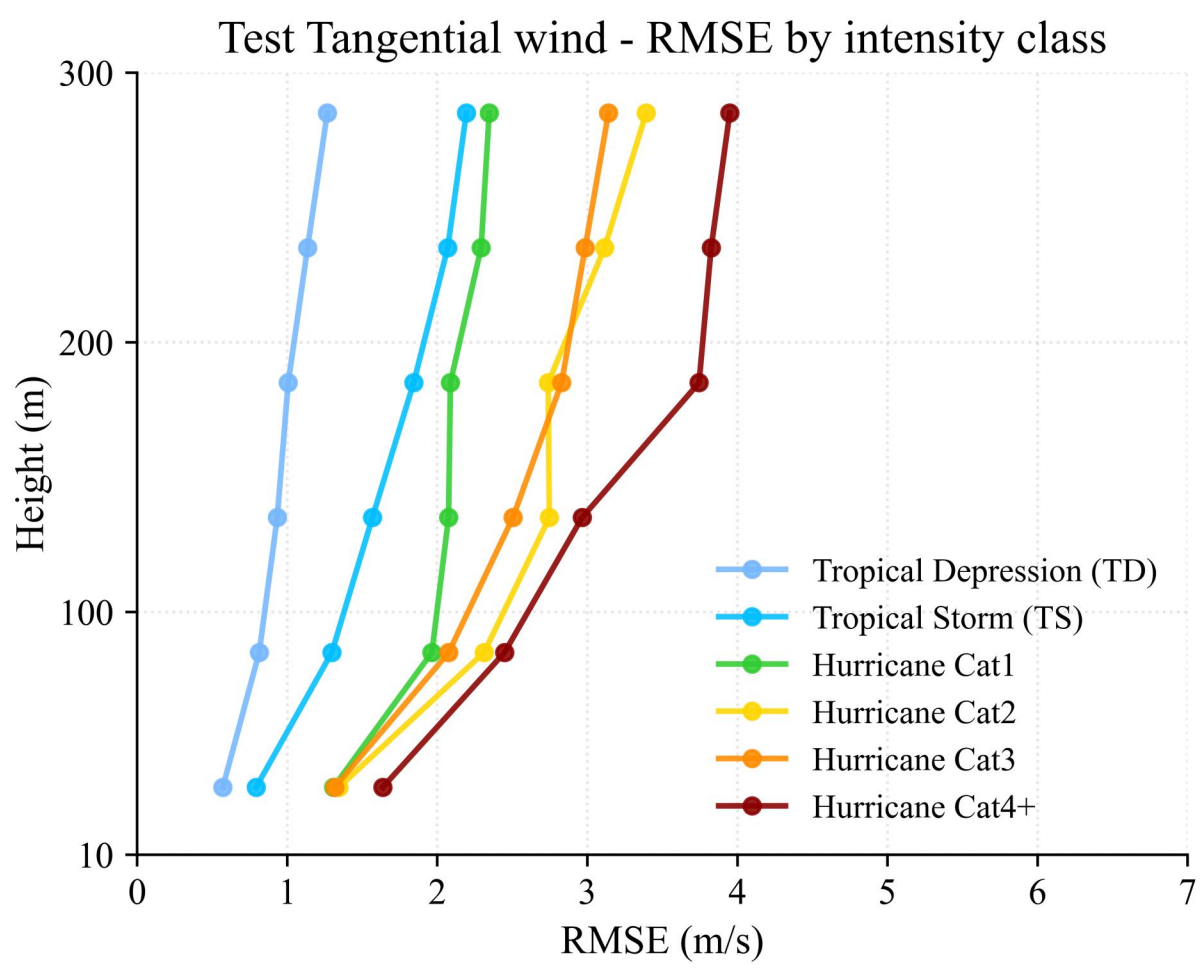

**Figure 12.** Height-dependent RMSE for tangential wind by intensity category

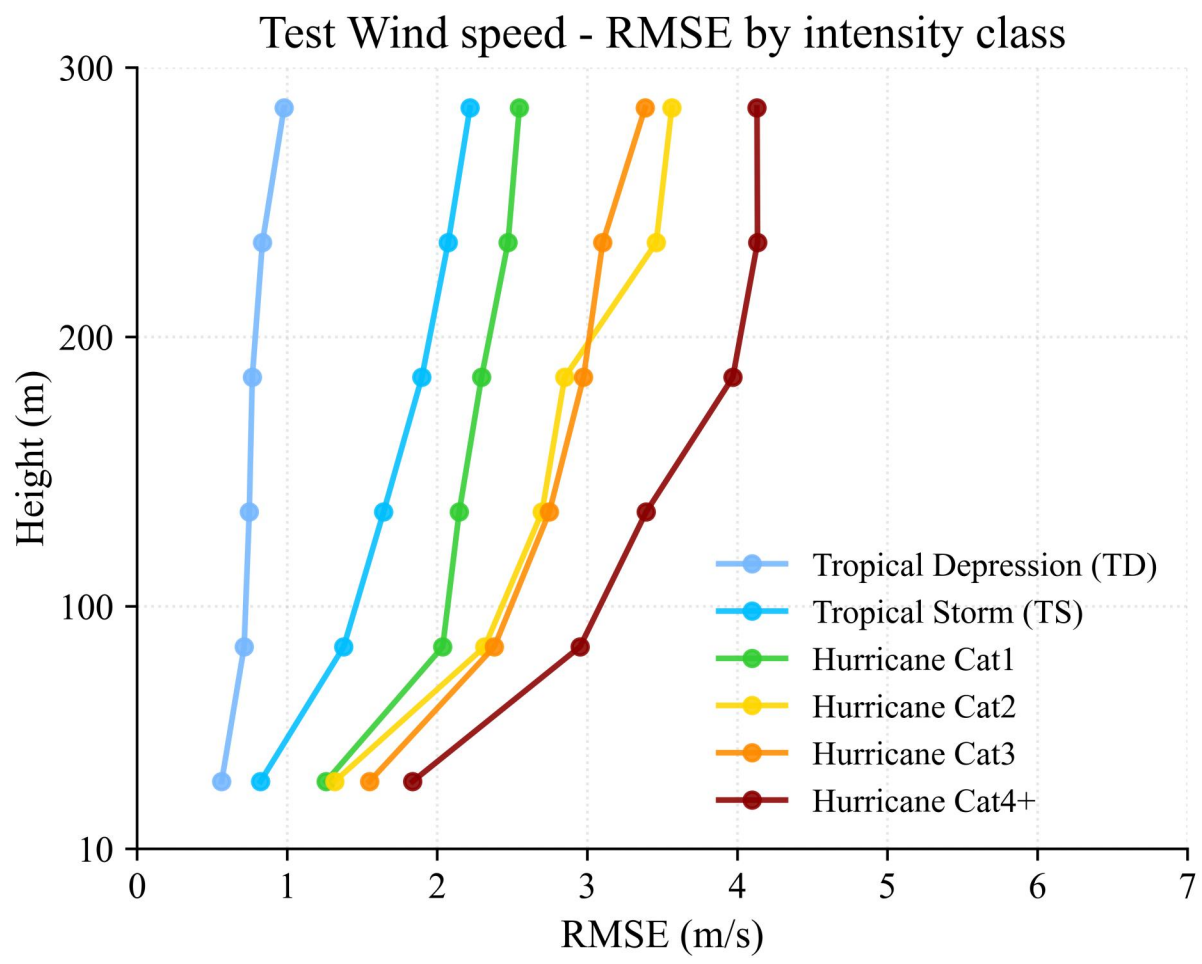

**Figure 13.** Height-dependent RMSE for wind-speed magnitude by intensity category

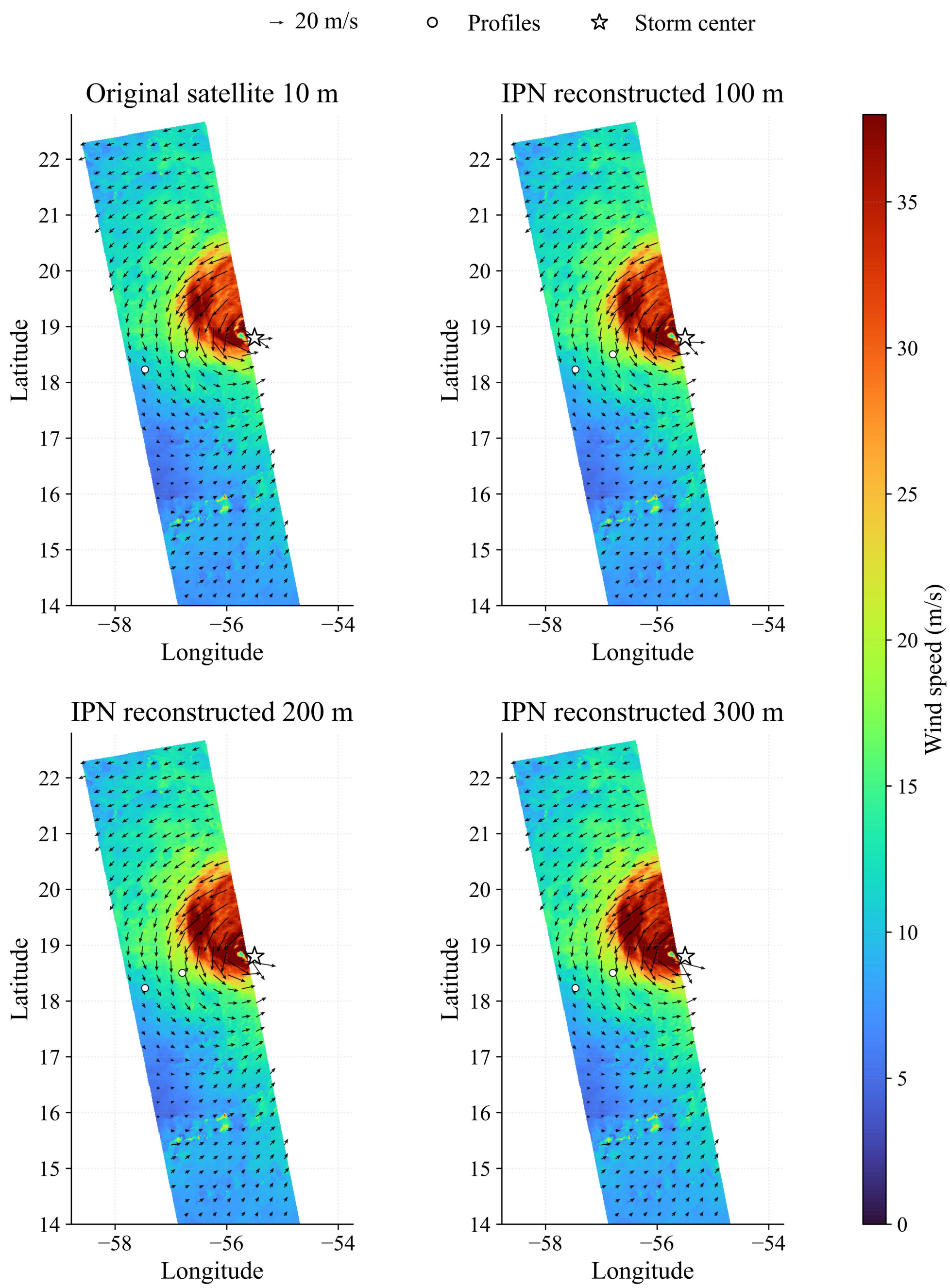

**Figure 14.** Spatial reconstruction of Hurricane Lee wind fields from SAR-derived 10 m winds.

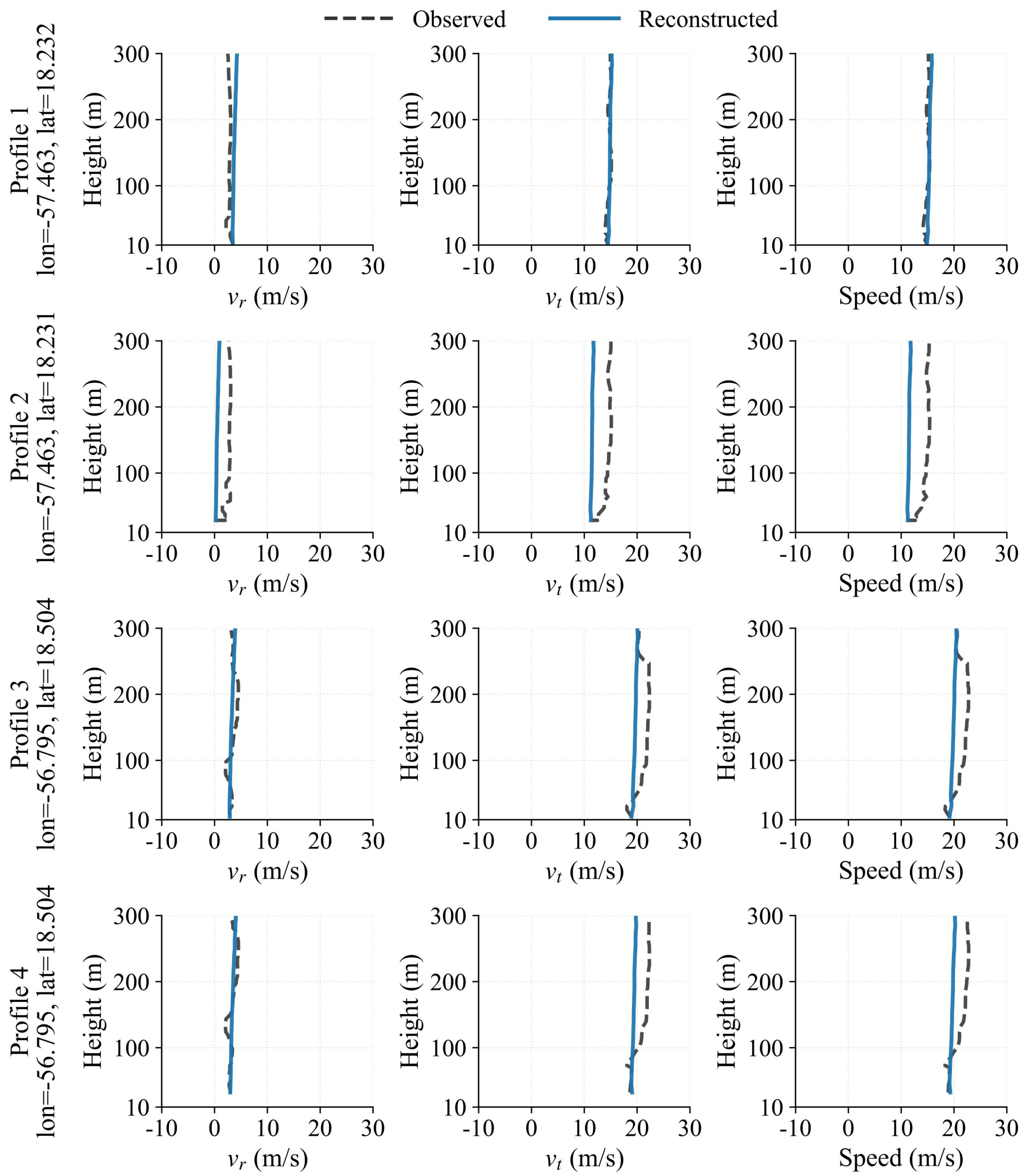

**Figure 15.** Dropsonde-based validation of Smartyphoon-BL3D reconstructed vertical profiles for Hurricane Lee.

Table 5. Profile-level reconstruction error statistics for the Hurricane Lee case

| Profile | $v_r$ MAE (m/s) | $v_r$ RMSE (m/s) | $v_t$ MAE (m/s) | $v_t$ RMSE (m/s) | Speed MAE (m/s) | Speed RMSE (m/s) |
|---|---|---|---|---|---|---|
| Profile 1 | 0.917 | 0.983 | 0.234 | 0.285 | 0.367 | 0.433 |
| Profile 2 | 2.133 | 2.187 | 2.937 | 3.025 | 3.175 | 3.269 |
| Profile 3 | 0.561 | 0.661 | 1.689 | 1.923 | 1.701 | 1.959 |
| Profile 4 | 0.456 | 0.566 | 1.863 | 2.093 | 1.852 | 2.085 |

### 4.6 Reconstruction Case

To further examine the applicability of the proposed Smartyphoon-BL3D to spatially continuous wind-field reconstruction, a case study was conducted using SAR-derived ocean-surface 10 m wind observations from Hurricane Lee in 2023, together with dropsonde profiles collected within the corresponding observation time window. The SAR 10 m wind field was used as the reference-height input, and the trained Smartyphoon-BL3D was applied to reconstruct wind fields at 100, 200, and 300 m. The reconstructed fields were then evaluated from two perspectives: spatial consistency with the original 10 m wind structure and profile-level agreement with available dropsonde observations.

For the dropsonde-based validation, sondes were required to have been released within ±1 h of the SAR overpass time to ensure that the storm environment remained comparable. Sonde drift was accounted for by using the GPS-reported latitude and longitude at each individual altitude level (specifically at 100, 200, and 300 m) rather than the release-point location, following the approach of Uhlhorn et al. (2014). Profiles that did not contain valid wind measurements at all three target heights were excluded from the comparison. Four dropsonde profiles met these criteria and were used for the quantitative evaluation.

As shown in Figure 14, the Smartyphoon-BL3D reconstructed wind fields at 100, 200, and 300 m preserve the main horizontal organization of Hurricane Lee observed in the original SAR 10 m wind field. The strong-wind region remains concentrated near the storm-center area, while the surrounding vector pattern maintains a coherent cyclonic structure. No obvious discontinuities or unrealistic spatial noise are introduced during the vertical reconstruction process. Meanwhile, as shown in Figure 15, the wind-speed distribution varies smoothly with height, indicating that the model does not simply replicate the 10 m field but introduces height-dependent modifications learned from dropsonde-based vertical-profile information. This result suggests that the differential anchoring layer and storm-relative conditional inputs provide a stable basis for extending satellite-derived surface winds to upper near-surface levels.

As shown in Table 5, averaged across the four validation profiles, the MAE and RMSE are 1.017 m/s and 1.099 m/s for radial wind, 1.681 m/s and 1.832 m/s for tangential wind, and 1.774 m/s and 1.936 m/s for wind-speed magnitude, respectively. These results are consistent with the previous test-set analysis: the model maintains strong reconstruction capability for general profile morphology, while larger errors occur in locally complex or dynamically heterogeneous regions. Therefore, the Hurricane Lee case demonstrates that the Smartyphoon-BL3D can transform satellite-derived 10 m winds into physically continuous low-level wind fields at engineering-relevant heights, while also maintaining reasonable agreement with independent dropsonde profiles. The Hurricane Lee case study demonstrates that the proposed framework generalizes beyond the statistical evaluation and can reconstruct physically realistic three-dimensional wind structures from independent observations.

### 4.7 Ablation Analysis of Physical Regularization

To evaluate the contribution of each physical regularization term, we conduct an ablation study with nine configurations. Because the finite-difference terms ($L_{\mathrm{curv}}$, $L_{\mathrm{slope}}$) produce raw values 4-5 orders of magnitude smaller than the direct wind-component terms at initialization, we calibrate the weights from a single forward pass of the initialized model: $\lambda_i = \frac{p_i}{p_{\mathrm{fit}}} \times \frac{L_{\mathrm{fit}}^{\mathrm{raw}}}{L_i^{\mathrm{raw}}}$, $\lambda_{\mathrm{fit}} = 1.0$. The calibrated weights are listed in Table 6.

Nine configurations are compared: a baseline using all seven constraints, a fit-only configuration with all constraints disabled, and seven single-removal configurations (no_curv through no_slope) each disabling one constraint while retaining the other six. All configurations share identical architecture, data split, optimizer, and training budget, differing only in loss composition.

**Table 6.** Physical constraints, calibrated weights, and raw loss values from an initialized model

| Loss term | Physical constraint | $\lambda$ | $L^{\text{raw}}$ |
|---|---|---|---|
| $L_{\text{fit}}$ | Adaptive Huber fidelity | 1.0 | 1.84E0 |
| $L_{\text{curv}}$ | Second-derivative smoothness | 8.21E3 | 1.76E-5 |
| $L_{\text{hemi}}$ | Hemisphere sign consistency | 7.35E-1 | 1.61E-1 |
| $L_{\text{angle}}$ | Inflow-angle sign and monotonicity | 1.12 | 1.18E-1 |
| $L_{\text{log}}$ | Log-law soft prior | 1.00E-1 | 1.05E0 |
| $L_{\text{film}}$ | FiLM modulation stability | 1.38E1 | 6.68E-3 |
| $L_{\text{int}}$ | Low-rank coupling L2 | 8.44E1 | 9.35E-4 |
| $L_{\text{slope}}$ | First-derivative smoothness | 7.65E2 | 1.55E-4 |

**Table 7.** Physics consistency metrics across ablation configurations

| Config | Hemi viol. | Inflow sign viol. | Log-law RMSE (m/s) | Curv excess $v_r$ | Curv excess $v_t$ | Slope mean |
|---|---|---|---|---|---|---|
| baseline | 0.0298 | 0.2401 | 1.19 | 0.00000 | 0.00000 | 0.00563 |
| fit_only | 0.0511 | 0.2550 | 1.29 | 0.00173 | 0.00383 | 0.00849 |
| no_curv | 0.0317 | 0.2407 | 1.17 | 0.00000 | 0.00000 | 0.00564 |
| no_hemi | 0.0516 | 0.2433 | 1.20 | 0.00000 | 0.00000 | 0.00538 |
| no_inflow | 0.0301 | 0.2521 | 1.19 | 0.00000 | 0.00000 | 0.00572 |
| no_log | 0.0315 | 0.2418 | 1.29 | 0.00000 | 0.00000 | 0.00565 |
| no_film | 0.0340 | 0.2441 | 1.21 | 0.00000 | 0.00000 | 0.00584 |
| no_coupling | 0.0319 | 0.2394 | 1.19 | 0.00000 | 0.00000 | 0.00558 |
| no_slope | 0.0342 | 0.2418 | 1.10 | 0.00000 | 0.00000 | 0.00863 |

**Table 8.** Marginal effect of removing each constraint ($\Delta$ from baseline, positive = degradation)

| Removed | $\Delta$ Hemi viol. | $\Delta$ Inflow sign viol. | $\Delta$ Log-law RMSE (m/s) | $\Delta$ Slope mean | Primary role |
|---|---|---|---|---|---|
| curv | +0.0019 | +0.0005 | -0.022 | +0.00001 | Curvature suppression |
| hemi | +0.0218 | +0.0032 | +0.013 | -0.00025 | Hemisphere compliance |
| inflow | +0.0003 | +0.0120 | -0.001 | +0.00010 | Inflow sign consistency |
| log | +0.0017 | +0.0016 | +0.104 | +0.00002 | Log-law profile shape |
| film | +0.0041 | +0.0039 | +0.025 | +0.00021 | Modulation stability |
| coupling | +0.0021 | -0.0008 | +0.006 | -0.00005 | Coupling stability |
| slope | +0.0044 | +0.0017 | -0.090 | +0.00301 | Profile smoothness |
| all (fit_only) | +0.0212 | +0.0149 | +0.107 | +0.00287 | — |

Several conclusions follow from the ablation results.

Physical consistency is evaluated on the test set using the following metrics: hemisphere violation rate (fraction of points where $\text{sign}_f \times v_t < 0$), inflow-angle sign violation rate, log-law RMSE (RMS deviation between predicted wind speed and the best-fit logarithmic profile within 10-40 m), curvature excess rate (fraction of second-derivative magnitudes exceeding a physically motivated threshold, computed separately for $v_r$ and $v_t$), and mean absolute first-order slope $|dp/dz|$. Table 7 presents results across all nine configurations; Table 8 reports the marginal effect of each constraint as $\Delta = \text{removed} - \text{baseline}$.

First, curvature regularization ($L_{\text{curv}}$) is the sole mechanism responsible for suppressing high-frequency second-derivative artifacts. Every configuration that retains $L_{\text{curv}}$ achieves a curvature excess rate of exactly zero for both $v_r$ and $v_t$, including no_slope which lacks first-order smoothing. Only the fully unconstrained fit_only configuration produces non-zero curvature excess ($v_r$: 0.00173, $v_t$: 0.00383), confirming that this penalty, rather than architectural inductive bias, eliminates spurious oscillations.

Second, the hemisphere sign consistency term ($L_{\text{hemi}}$) is the dominant constraint enforcing the correct cyclonic circulation direction. Removing $L_{\text{hemi}}$ raises the hemisphere violation rate from 0.0298 to 0.0516, essentially matching fit_only (0.0511). All other single-removal configurations remain within 0.030-0.034, demonstrating that the $\text{sign}_f$ encoding in the conditioning vector alone is insufficient to guarantee hemispheric consistency without the explicit soft constraint.

Third, the first-order slope regularization ($L_{\text{slope}}$) and the log-law soft prior ($L_{\text{log}}$) each control distinct aspects of the vertical profile shape. Removing $L_{\text{slope}}$ increases the mean absolute slope from 0.00563 to 0.00863 (fit_only: 0.00849), while all other single-removal configurations stay within 0.0054-0.0058, confirming that this term is the primary roughness control. Removing $L_{\text{log}}$ increases the log-law RMSE from 1.19 to 1.29 m/s (fit_only: 1.29 m/s), indicating that the constraint effectively guides the near-surface wind profile toward the classical logarithmic form.

Fourth, the inflow-angle constraint ($L_{\text{angle}}$) provides a measurable reduction in inflow sign violations (0.2401 to 0.2521 when removed). The FiLM stability ($L_{\text{film}}$) and low-rank coupling ($L_{\text{int}}$) terms produce smaller effects on the 30-epoch timescale examined here; this is consistent with their intended role as training-stability guards that prevent unbounded growth of modulation parameters and coupling coefficients during prolonged optimization, rather than directly shaping the instantaneous wind profile morphology.

In summary, the ablation study confirms that each of the seven physical regularization terms addresses a specific and non-redundant aspect of physical consistency. The calibration strategy provides equitable initial gradient contributions despite the large disparity in raw loss magnitudes, enabling a controlled comparison that isolates the marginal effect of each constraint. Overall, the results demonstrate that the proposed physics-constrained implicit representation accurately reconstructs continuous tropical cyclone near-surface wind profiles while preserving key characteristics of boundary-layer dynamics across diverse storm conditions. The model is accurate because it successfully reconstructs the key dynamical characteristics of the tropical cyclone boundary layer, including radial inflow, tangential wind evolution, vertical shear, and inflow-angle structure.

## 5 Discussion

### 5.1 Physical Interpretation of the Reconstructed Wind Profiles

The results demonstrate that the proposed Physics-Constrained Implicit Profile Network (Smartyphoon-BL3D) successfully reconstructs the principal characteristics of tropical cyclone near-surface wind profiles from sparse dropsonde observations while maintaining physically realistic boundary-layer structure. Unlike conventional interpolation methods that estimate winds only at discrete heights, the continuous implicit representation captures the smooth vertical evolution of radial inflow, tangential circulation, and wind direction throughout the 10–300 m layer. The close agreement between reconstructed and observed profiles suggests that the model has learned physically meaningful relationships between storm-scale environmental conditions and boundary-layer wind structure rather than simply interpolating between measurements.

The reconstructed profiles preserve several well-established characteristics of tropical cyclone boundary layers, including the gradual increase of tangential wind with height, the turning of the wind vector toward the storm center near the surface, and the continuous variation of inflow angle. These features are fundamental to frictionally driven boundary-layer dynamics and directly influence air–sea momentum exchange, surface stress, and storm intensity evolution. The ability to reproduce these structures across storms of different intensities indicates that the proposed framework captures the dominant physical controls governing near-surface wind-profile variability.

### 5.2 Role of Physics-Constrained Learning

A distinguishing feature of the proposed framework is the integration of physical knowledge into a continuous neural representation. Tropical cyclone observations are inherently sparse, irregularly distributed, and subject to observational uncertainty, making purely data-driven models susceptible to unrealistic oscillations, unstable extrapolation, and physically inconsistent profile evolution. By incorporating physically motivated constraints during training, the proposed framework guides the optimization toward solutions that remain consistent with established boundary-layer behavior while preserving the flexibility of deep learning.

Rather than explicitly solving the governing equations of atmospheric motion, the model incorporates prior physical knowledge through constraints that preserve observed surface winds, encourage smooth vertical profile evolution, maintain realistic inflow-angle behavior, and improve optimization stability. These constraints reduce physically implausible solutions while allowing the network to learn complex nonlinear relationships directly from observations. The results demonstrate that combining continuous implicit neural representations with physically informed regularization provides an effective balance between data fidelity and physical consistency, illustrating the broader value of hybrid physics–AI approaches for atmospheric reconstruction problems.

### 5.3 Comparison with Previous Approaches

The reconstruction of tropical cyclone near-surface wind profiles has traditionally relied on analytical or empirical formulations derived from boundary-layer theory and field observations. Classical logarithmic-law, power-law, and Ekman-type models provide physically interpretable descriptions of the vertical wind structure and have been widely used in boundary-layer research, engineering design, and operational applications (Kepert, 2001; Smith and Montgomery, 2010; Zhang and Uhlhorn, 2012). However, these approaches generally assume prescribed functional forms with a limited number of free parameters, limiting their ability to represent the complex, nonlinear, and storm-dependent wind structures observed in intense tropical cyclones. Rapid changes in vertical shear, strong inflow-angle variations, and localized profile curvature often depart substantially from these idealized formulations.

A second class of approaches employs interpolation, statistical regression, or conventional machine-learning methods to estimate winds at discrete heights from sparse observations. Although these methods often improve reconstruction accuracy relative to empirical formulations, they generally predict winds only at predefined vertical levels and frequently require post-processing to generate continuous profiles. Moreover, because physical constraints are not explicitly incorporated, reconstructed profiles may exhibit unrealistic oscillations or inconsistencies with known boundary-layer behavior, particularly when observations are sparse or noisy.

Recent advances in artificial intelligence have demonstrated the potential of deep learning for reconstructing complex atmospheric fields from incomplete observations. Increasingly, Earth-system applications have combined machine learning with physical knowledge to improve robustness, generalization, and physical consistency (e.g., Reichstein et al., 2019). The present framework follows this emerging paradigm by integrating a continuous implicit neural representation with physically informed constraints that preserve observed surface winds while promoting realistic vertical wind-profile evolution. Rather than prescribing an analytical profile shape, the model learns a continuous mapping between storm conditions, height, and wind components directly from observations. This hybrid approach combines the flexibility of deep learning with established understanding of tropical cyclone boundary-layer dynamics, providing a more physically consistent reconstruction framework than either traditional empirical formulations or unconstrained data-driven models.

### 5.4 Applications and Broader Implications

The capability to reconstruct continuous near-surface wind profiles from sparse observations has potential applications across both tropical cyclone research and operational forecasting. Continuous vertical wind profiles provide a more complete description of radial inflow, tangential circulation, vertical shear, and inflow-angle evolution than isolated measurements at discrete heights, enabling more comprehensive investigations of air–sea interaction, boundary-layer dynamics, and tropical cyclone intensity change.

The proposed framework also enhances the value of existing and future observing systems. By extending sparse aircraft or satellite observations into continuous three-dimensional near-surface wind fields while preserving observed surface conditions, the model offers a practical tool for validating numerical weather prediction models, improving model initialization, and supporting future data assimilation systems that assimilate heterogeneous atmospheric observations. Because the framework predicts continuous profiles rather than values at discrete levels, it can readily accommodate measurements from multiple observing platforms with different vertical sampling characteristics.

Beyond atmospheric science, accurate reconstruction of near-surface wind profiles has important applications in offshore engineering and coastal hazard assessment. Continuous wind profiles provide essential information for estimating wind loads on offshore platforms, floating wind turbines, ships, and coastal infrastructure, as well as for improving storm-surge and wave models that depend on the vertical distribution of wind stress. More broadly, this study illustrates how physics-constrained artificial intelligence can bridge sparse atmospheric observations and physically consistent environmental reconstructions. As Earth-system observing networks continue to expand, similar hybrid AI frameworks may become increasingly valuable for reconstructing continuous environmental fields and supporting next-generation Earth-system prediction systems.

### 5.5 Limitations and Future Work

Several limitations of this study should be acknowledged. First, the model is trained using NOAA Hurricane Research Division dropsonde observations collected primarily over the Atlantic and eastern North Pacific, and its performance in other ocean basins and environmental regimes remains to be evaluated. Second, the present framework reconstructs wind profiles only within the observed near-surface layer (10–300 m) and does not explicitly represent the full depth of the tropical cyclone boundary layer. Third, although physically informed constraints substantially improve reconstruction realism, these constraints represent empirical approximations of established boundary-layer behavior rather than direct enforcement of the governing equations. Future work may further improve the framework by incorporating additional observations from Doppler radar, airborne lidar, uncrewed aircraft systems, and satellite platforms, as well as by integrating the reconstruction framework with numerical weather prediction, data assimilation, and operational forecasting systems.

## 6 Conclusions

This study developed a Physics-Constrained Implicit Profile Network (Smartyphoon-BL3D) for reconstructing continuous tropical cyclone near-surface radial and tangential wind profiles from sparse dropsonde observations and storm-environmental information. By combining continuous implicit neural representations with physically informed constraints, the proposed framework reconstructs wind profiles while preserving observed 10 m wind conditions and promoting physically realistic vertical wind-profile evolution. The framework addresses the limitations of conventional empirical profile models and discrete interpolation methods by providing continuous wind estimates throughout the near-surface boundary layer.

Evaluation using quality-controlled NOAA Hurricane Research Division dropsonde observations collected from 93 tropical cyclones during 2001–2022 demonstrates that the proposed framework accurately reconstructs the vertical structure of tropical cyclone near-surface winds across a broad range of storm intensities. The reconstructed profiles reproduce key characteristics of tropical cyclone boundary-layer flow, including radial inflow, tangential wind evolution, vertical shear, and inflow-angle variation, indicating that the model captures physically meaningful relationships between storm conditions and boundary-layer structure. The Hurricane Lee (2023) case study further illustrates the capability of the framework to extend sparse observations into continuous three-dimensional near-surface wind fields while maintaining consistency with independent measurements.

Beyond reconstruction accuracy, this study demonstrates the value of integrating physical knowledge with modern machine-learning techniques for atmospheric applications. Rather than relying solely on data-driven learning, the incorporation of physically motivated constraints improves reconstruction realism and provides a flexible framework that remains consistent with established boundary-layer behavior. The results highlight the potential of hybrid physics–AI approaches to bridge sparse atmospheric observations and physically consistent environmental reconstructions.

The proposed framework has potential applications in tropical cyclone boundary-layer research, satellite surface-wind extension, numerical weather prediction, data assimilation, offshore engineering, storm-surge and wave modeling, and coastal hazard assessment. As atmospheric observing systems continue to expand, physics-constrained continuous reconstruction methods may provide an increasingly valuable component of future Earth-system analysis and prediction systems.

Future work will extend the framework to additional observational datasets, broader environmental conditions, and other ocean basins, while exploring integration with numerical weather prediction and operational forecasting systems. More broadly, this study demonstrates how physics-constrained artificial intelligence can enhance the interpretation of sparse atmospheric observations and contribute to the next generation of Earth-system science and environmental prediction.

## Data Availability

The observational data supporting this study come from the NOAA/AOML Hurricane Research Division dropsonde archive, spanning 2001 through 2022 (https://www.aoml.noaa.gov/hrd/data_sub/dropsonde.html). After quality control — removal of profiles with missing or physically implausible wind measurements, a Hampel-type outlier filter, and a requirement of at least three valid points in each 100-meter layer — we retained 7,524 profiles from 93 distinct tropical cyclones. The quality-controlled dataset is archived at MA, J., YANG, Y., ROGERS, R., ZHANG, J. A.& TANG, J. (2026). A Physics-Constrained Implicit Profile Network for Continuous Reconstruction of Tropical Cyclone Near-Surface Wind Profiles [Dataset]. Zenodo. https://doi.org/10.5281/zenodo.21868265.

The Sentinel-1 SAR-derived ocean-surface wind field for Hurricane Lee (2023), used in the reconstruction case study, was obtained from the Copernicus Open Access Hub (https://scihub.copernicus.eu/).

## Conflict of Interest

The authors declare there are no conflicts of interest for this manuscript.

## Author CRediT

Jian Ma led the conceptualization, methodology design, software implementation, formal analysis, and visualization, and contributed to writing the original draft and revising the manuscript.

Yilin Yang contributed to methodology development, model implementation, investigation, and manuscript writing and revision.

Robert Rogers provided the HRD dropsonde data, contributed expert guidance on tropical cyclone boundary layer dynamics, and critically reviewed and revised the manuscript.

Jie Tang co-conceptualized the study, contributed to validation and investigation, supervised the research, acquired funding, managed the project, and critically reviewed and revised the manuscript.

Jun A. Zhang provided dropsonde data expertise, contributed to validation, supervised the boundary-layer interpretation, acquired funding, and critically reviewed and revised the manuscript.

## Inclusion in Global Research Statement

This research represents an international collaboration between the Asia-Pacific Typhoon Collaborative Research Center (Shanghai, China), the Shanghai Typhoon Institute of the China Meteorological Administration (Shanghai, China), and the University of Miami (Miami, FL, USA), in partnership with NOAA's Hurricane Research Division (Miami, FL, USA).

Research Collaboration Disclosure:

1. The HRD dropsonde data used in this study are publicly available through the NOAA/AOML Hurricane Research Division archive. No special permits, authorizations, or formal agreements with local communities were required for data access or use.
2. The computing resources used for model training were provided by the Asia-Pacific Typhoon Collaborative Research Center computing infrastructure.
3. All co-authors from all participating institutions meet AGU Publications authorship criteria and were substantively involved in conceptualization, methodology, data curation, analysis, and/or manuscript preparation throughout the research process.
4. No local collaborators were excluded from authorship; all individuals who contributed substantively to this work are listed as co-authors.

End Users of the Research:

1. The primary end users of this research include:
2. Tropical cyclone forecasters and researchers seeking improved boundary-layer wind-profile characterization
3. Offshore structural and wind-energy engineers requiring near-surface wind-profile data for design standards

4. Storm-surge modelers who use wind-profile inputs for coastal hazard assessment
5. Coastal risk assessment and emergency management agencies
6. The broader tropical cyclone boundary-layer dynamics and air – sea interaction research communities

Ethical Considerations: This research was conducted in accordance with the TRUST Code of Conduct for Equitable Research Partnerships. All data are publicly available and results are openly shared. The authors acknowledge the ongoing support and collaboration of the NOAA/AOML Hurricane Research Division, whose dropsonde archive has enabled decades of tropical cyclone boundary-layer research.

**Acknowledgments**

Funding: The study was supported by the Science and Technology Project of State Grid Corporation of China(521104250028-234-TD) and the ESCAP/WMO Project (EXOTICCA) , Typhoon Scientific and Technological Innovation Group of China Meteorological Administration CMA2023ZD06）

Data and Technical Acknowledgments: The authors gratefully acknowledge the NOAA/AOML Hurricane Research Division (HRD) for collecting, quality-controlling, and publicly archiving the dropsonde observations that made this study possible. We thank the flight crews and scientists of the NOAA Hurricane Hunters and the U.S. Air Force Reserve 53rd Weather Reconnaissance Squadron for their dedication in obtaining these valuable in-situ observations under extreme conditions. The Sentinel-1 SAR wind data for the Hurricane Lee case study were obtained from the Copernicus Marine Service / European Space Agency.